\documentclass[12pt]{article}
\usepackage[dvips]{graphicx}
\usepackage{amsmath}
\usepackage{graphicx}
\usepackage{amsfonts}
\usepackage{amssymb}
\usepackage{epsfig}
\usepackage{subfigure}
\usepackage[usenames]{color}

\newcommand{\be}{\begin{equation}}
\newcommand{\ee}{\end{equation}}
\newcommand{\bea}{\begin{eqnarray}}
\newcommand{\eea}{\end{eqnarray}}
\newcommand{\ba}{\begin{array}}
\newcommand{\ea}{\end{array}}
\newcommand{\beas}{\begin{eqnarray*}}
\newcommand{\eeas}{\end{eqnarray*}}
\newcommand{\bes}{\begin{equation*}}
\newcommand{\ees}{\end{equation*}}

\def\i2           {\mbox{$\frac{i}{2}$}}

\begin{document}
\title{\bf Exotic Multi-quark States in the Deconfined Phase from Gravity Dual Models}

\author{P. Burikham\thanks{Email:piyabut@gmail.com}, A. Chatrabhuti\thanks{Email:auttakit.c@chula.ac.th}, E. Hirunsirisawat\thanks{Email:ekapong.hirun@gmail.com}\\
{\small {\em Theoretical High-Energy Physics and Cosmology Group,
Department of Physics,}}\\ {\small {\em Faculty of Science,
Chulalongkorn University, Bangkok 10330, Thailand}}}

\maketitle

\begin{abstract}
\noindent  \\
In the deconfined phase of quark-gluon plasma, it seems that most
of the quarks, antiquarks and gluons should be effectively free in
the absence of the linear confining potential.  However, the
remaining Coulomb-type potential between quarks in the plasma
could still be sufficiently strong that certain bound states,
notably of heavy quarks such as $J/\psi$ are stable even in the
deconfined plasma up to a certain temperature.  Baryons can also
exist in the deconfined phase provided that the density is
sufficiently large.
 We study three kinds of exotic multi-quark bound states in the
deconfined phase of quark-gluon plasma from gravity dual models in
addition to the normal baryon. They are $k$-baryon,
$(N+\bar{k})$-baryon and a bound state of $j$ mesons which we call
``$j$-mesonance". Binding energies and screening lengths of these
exotic states are studied and are found to have similar properties
to those of mesons and baryons at the leading order. Phase diagram
for the exotic nuclear phases is subsequently studied in the
Sakai-Sugimoto model.  Even though the exotics are less stable
than normal baryons, in the region of high chemical potential and
low temperature, they are more stable thermodynamically than the
vacuum and chiral-symmetric quark-gluon plasma
phases~($\chi$S-QGP).

\end{abstract}

\newpage
\section{Introduction}

The discovery of AdS/CFT correspondence \cite{maldacena, agmoo}
provides a new tool for studying the strongly coupled gauge
theories.  Although the original setup and most of the systems
that string theorists have been investigating so far are highly
supersymmetric and conformal, a lot of progress has been made in
constructing more realistic models.  Now we have examples of
QCD-like gauge theory with known gravity dual that share most of
the qualitative features of QCD.  These holographic models allow
us to perform analytic calculations in the regimes which are too
difficult to implement for the real QCD even for lattice
calculations.  The properties of quark-gluon plasma from
Relativistic Heavy Ion Collisions and QCD at finite baryon density
are two examples of such regimes.

The gravity dual of baryons can be described via baryon vertex
\cite{witb, gross&ooguri}, a D-brane wrapping higher dimensional
sphere in 10-dimensional curved background with $N$ strings
attached to it and ending at the boundary.  These strings are
required to cancel an $N$ charge in the world-volume of the
wrapped brane due to the presence of RR flux in the background.
The endpoint of fundamental string that ends on D-brane is
electrically charged with respect to world-volume $U(1)$ gauge
field.  ItÕs charge is $+1$ or $-1$ depending on the orientation
of the string and D-brane. Moreover, strings stretching from the
baryon vertex to the boundary of $AdS$ or the corresponding
background spacetime~(e.g. in Sakai-Sugimoto model) behave as
fermions, giving antisymmetricity to the baryon vertex. This fact
allows us to construct an $SU(N)$ gauge-invariant combination of
$N$ quarks as required by the group theory.  Baryon configurations
were investigated further in \cite{Imamura}-\cite{Guijosa}.  The
authors in \cite{BISY} extended the consideration in confining
background where it was found that the binding energy is linear in
$N$ and in the ÒsizeÓ of the baryon on the boundary.  And
furthermore, they found that in $\mathcal{N}_{SUSY} =4$ theory
there are stable configurations for baryons which are made of $k$
quarks, or ``k-baryon", if $5N/8<k\leq N$.  Such configurations
can be realized by considering the usual baryon vertex with $k$
strings stretched up to the boundary and the rest $N-k$ strings
stretched down to the horizon. These baryons are not colour
singlet and transform as $\frac{N!}{k!(N-k)!}$ representation
under $SU(N)$ gauge group, for example the case $k = N-1$ gives
rise to a baryonic configuration in the anti-fundamental
representation.  In a confining theory we do not expect to find
such a bound state. It was proposed in \cite{gho_multi-q,
gho_k-quark} that the $k<N$ bound states can exist in a deconfined
phase.

 In general, we could imagine that there would be more exotic baryon
states in the deconfined phase where bound states of quarks need
not be the colour singlet.  Some attempts have been made in
constructing holographic description of exotic multi-quarks bound
states\cite{gho_multi-q}-\cite{Wen}.  The author in \cite{Wen}
considered exotic quark configurations formed by combining two or
more baryon vertices together.  However, it might be possible to
construct an exotic baryon from a single baryon vertex which
should be more energetically preferable.  One useful observation
is that there are infinite combinations of string charges that can
cancel the charge from the background RR flux.  Hence, the total
number of strings attached to the baryon vertex need not to be
equal to $N$. For example, if the orientation of D-branes is fixed
in such a way that there is $+N$ units of charge on its
world-volume, we can attach $N+k$ strings, each with $-1$ charge
and $k$ strings with $+1$ charge to make the total charge
vanishes. As long as the conservation of charge is concerned, $k$
could be any integer. In this case, we can construct a $k > N$
baryon. Such baryon could be the lightest bound state in some
irreducible representation of the underlying gauge theory thus it
may be stable and can be observed in the deconfined phase.  We
would like to investigate this possibility further in this paper.

It is also interesting to study exotic baryons in more realistic
model such as Sakai-Sugimoto model \cite{ss lowE, ss more}. This
model is based on Witten's model \cite{Wittcon} which uses the
D4-brane wrapping a Scherk-Schwarz circle and adds a stack of $N_f
$ probe D8-branes and a stack of $N_f$ probe anti-D8-branes
transverse to the circle.  This model contains massless chiral
fermions and the flavour symmetry. The most striking feature of
this model is that it introduces geometrical mechanism for
spontaneous chiral symmetry breaking. Using the fact that the
circle vanishes at a finite radial coordinates in the near horizon
limit, D8-branes and anti-D8-branes are connected in a U-shaped
configuration.  At low temperature the model describes a confining
gauge theory with broken chiral symmetry. Above a deconfinement
temperature, gluons become effectively free.  However, both the
connected U-shape D8-branes configuration and the separated
parallel brane-anti-brane configuration are possible in the
intermediate temperature.  The chiral symmetry is still broken
even though the gluons are already deconfined.  At higher
temperature the chiral symmetry is restored, which is illustrated
geometrically by the separation of the D8-branes and
anti-D8-branes \cite{Aharony}.
 This corresponds to the branes being in parallel configuration.

The model also has an interesting phase structure.  Finite baryon
density in the Sakai-Sugimoto model has been studied in
\cite{Kim,Tanii} and extended to the full parameter space in
\cite{bll} where baryon matter is represented by D4-branes in the
D8-brane (nuclear matters) or by strings stretched from the
D8-brane down to the horizon (quark matters). It was shown that
the former configuration is always preferred to the latter and
quark matters are unstable to density fluctuations.  In the
deconfined phase there are three regions corresponding to the
vacuum, quark-gluon plasma, and nuclear matter, with a first-order
and a second-order phase transition separating these three phases.
The author in \cite{bll} found that for a large baryon number
density, and at low temperatures, the dominant phase has broken
chiral symmetry in agreement with QCD at high density. It is
interesting to see how exotic baryon states fit into the phase
structure.

This paper is organized as the following.  In section 2, we
discuss some classes of exotic baryon configurations and
investigate their static configurations in section 3. Binding
energy and screening length of the configurations are calculated
in section 4. The dependence on free quark mass of exotic baryon
configuration is discussed in section 5.  The phase diagram of
Sakai-Sukimoto model with exotic baryons is investigated in
section 6.  We discuss our results in section 7 and conclude in
section 8.

\section{Some classes of multi-quark states }

In the deconfined phase of QGP, coloured states of a number of
quarks and antiquarks can exist in the medium as long as it is
energetically more favoured than the free quarks and antiquarks or
other mesonic states.  We will call these multi-quark states as
``baryons" in this article.  In the confined phase, the only
allowed baryons are those with colour singlet combinations such as
nucleons and pentaquarks.  For the deconfined phase, baryons can
have colour and thus can have more varieties than the situation in
the confined phase.

In general, a D$(8-p)$-brane wrapping the subspace ${S}_{8-p}$ of
the background spacetime sources the gauge field $A_{(1)}$ on its
world volume. This gauge field will couple with the antisymmetric
$(8-p)$-form field strength $G_{(8-p)}$ and induce the charge upon
the wrapping D$(8-p)$-brane. If the background is generated by a
stack of $N$ D$p$-branes, then the charge being induced upon the
wrapping D$(8-p)$-brane will be exactly $N$. This charge needs to
be cancelled by external charges brought about by strings. Each of
the strings stretching out from the wrapping brane to the
spacetime boundary or probe branes carries $-1$ unit of charge.
Therefore it is required that the total number of ``quark" strings
stretching out from the wrapping brane must be $N$. The
configuration of wrapping D$(8-p)$-brane with totally $N$ strings
stretching out is called a baryon vertex~\cite{witb,gross&ooguri}.

For the confined phase, since quarks cannot exist as free-quark
strings with one end falling behind the horizon, therefore they
can only start from the baryon vertex and go to the probe branes.
On the other hand, in the deconfined phase, a radial string
configuration lying along the radial coordinate is also a
classical solution of the Nambu-Goto action~\cite{abl} and it
represents the free (anti)quark state in the QGP medium.  A string
can either start from the baryon vertex and go radially to the
horizon of the background spacetime or it can come from the
horizon and end at the baryon vertex.  We will call this string
configuration which is allowed in the deconfined phase as the
``radial string".

In the deconfined phase of QGP, it is possible to have
$k_\text{h}$ strings hanging from the spacetime boundary down to
the baryon vertex and another $k_\text{r}$ strings stretching
radially from the baryon vertex down to the horizon.  The total
number $k_\text{h}+k_\text{r} = N$ is the charge conservation
constraint on the configuration.  This configuration is known as
``{\bf $k$-baryon}"~\cite{BISY}.

\begin{figure}
\centering
\includegraphics[width=1.0\textwidth]{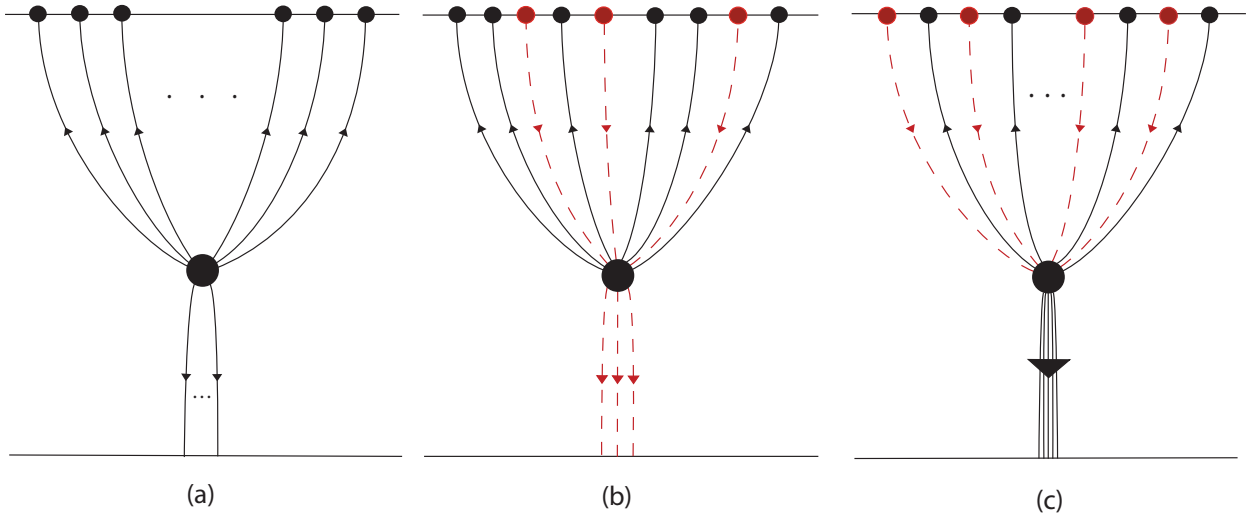} \caption{The gravity dual configurations of the hypothetical exotic
states (a) $k$-baryon with the number of hanging strings $k_{\text
h}=k<N$ and the number of radial strings $k_{\text r}=N-k$. (b)
$(N+ \bar{k})$-baryon with $k_{\text h}=N+\bar{k}$ and $k_{\text
r}=\bar{k}$. (c) $j$-mesonance with $k_{\text h}=2j$ and $k_{\text
r}=N$.} \label{exo}
\end{figure}

Another possible configuration is when there are $N$ quark-strings
and $\bar{k}$ antiquark-strings hanging down to the vertex from
the probe branes. To conserve the charge, there are additional $k$
quark-strings hanging from the vertex down to the horizon.  We
will call this configuration ``{\bf $(N+\bar{k})$-baryon}" (e.g.
pentaquark could be represented by one of this kind).

An even more interesting configuration allowed in the deconfined
phase is when there are $j$ pairs of quark and antiquark strings
hanging from the probe branes down to the vertex. Again, to
conserve charges, we need $N$ radial strings stretching from the
vertex down to the horizon.  This configuration obviously can
decay into $j$ mesons when it is less energetically favoured.
Therefore we will call this state, a ``{\bf $j$-mesonance}",
representing a binding state of $j$ mesons in the QGP.

In summary, the charge conservation constraint for each case can
be expressed as the following. \\

\noindent For $k$-baryon,
\begin{equation}\label{k-b_khkr}
 k_\text{h} + k_\text{r} =N; \quad  k_{\text{h}}=k.\\
\end{equation}

\noindent For ($N+\bar{k}$)-baryon,
\begin{equation}\label{N+kbar_khkr}
 k_\text{h} -k_\text{r} = N; \quad  k_{\text{h}}=N+\bar{k}.\\
\end{equation}

\noindent For $j$-mesonance,
\begin{equation}\label{j-mkhkr}
 k_\text{h}= 2 j; \qquad k_\text{r} =N.\\
\end{equation}
Note that $k_{\text{h}}$ is the number of strings hanging from the
boundary down to the baryon vertex and $k_{\text{r}}$ is the
number of strings hanging from the vertex down to the horizon. The
value of $\bar{k}$ and $j$ can be as large as $N\times N_{f}$.
However, in this article, we will take this number to be large and
ignore the upper bound on $\bar{k}$ and $j$.  Each configuration
of exotic baryons is illustrated in Fig.~\ref{exo}.

%\begin{figure}[htp]
%\centering
%\includegraphics[width=0.7\textwidth]{}
%\caption[]{ }\label{}
%\end{figure}

\section{Force conditions}

In this section, we will consider the force condition for each
exotic configuration of the quarks and antiquarks in a deconfined
phase. The calculation will be performed in the gravity background
similar to those of Sakai and Sugimoto's~\cite{ss lowE}. Even
though the chiral symmetry restoration can be addressed within
this model, we will not consider the aspect in this section but
rather focus our attention on the high temperature phase where
quarks and antiquarks are effectively free in the absence of the
linear confining potential. The positions of
D8/$\overline{\text{D8}}$ will be taken to be large and we will
approximate it to be infinity in this section as well as in the
discussion of binding energy and screening length in section
\ref{energy}.  Analysis in this heavy-quark limit provides us with
valuable physical understanding of certain essential features of
the exotic states.
 Generalized results for a near-horizon background metric of the
D$p$-branes solution and its dependence on positions of the probe
branes will be given in section \ref{mass dependence}.

Even in the deconfined phase, quarks and antiquarks feel effective
(screened) potential from other constituents. Therefore, a number
of population of them will exist in various forms of bound states,
some of which are exotic in the sense that they cannot be formed
in the confined phase at low temperature.

Start with the following background metric

\begin{equation}
ds^2=\left( \frac{u}{R_{D4}}\right)^{3/2}\left( f(u) dt^2 +
\delta_{ij} dx^{i}
dx^{j}+{dx_4}^2\right)+\left(\frac{R_{D4}}{u}\right)^{3/2}\left(u^2
d\Omega_4^2 + \frac{du^2}{f(u)}\right)\\ \nonumber
\end{equation}

\begin{equation}
F_{(4)}=\frac{2\pi N}{V_4} {\epsilon}_4, \quad \quad e^{\phi}=g_s
\left( \frac{u}{R_{D4}}\right)^{3/4} ,\quad\quad R_{D4}^3\equiv
\pi g_s N l_{s}^3,\nonumber
\end{equation}

\noindent where $f(u)\equiv 1-u_{T}^{3}/u^3$, $u_T=16{\pi}^2
R_{\text{D4}}^3 {T^2} /9$.  Note that the compactified $x_4$
coordinate ($x^4$ transverse to the probe D8 branes), with
arbitrary periodicity $2\pi R$, never shrinks to zero.  The volume
of the unit four-sphere $\Omega_4$ is denoted by $V_4$ and the
corresponding volume 4-form by $\epsilon_4$. $F_{(4)}$ is the
4-form field strength, $l_s$ is the string length and $g_s$ is the
string coupling.  The dilaton in this background has
$u$-dependence and its value changes along the radial direction
$u$. This is a crucial difference in comparison to the
AdS-Schwarzschild metric case where dilaton contribution is
constant.

The action of the baryon configuration is given by
\begin{equation}
S=S_{\text{D4}}+k_{\text{h}} S_{\text{F1}}+k_{\text{r}}
\tilde{S}_{\text{F1}},\label{total_action}
\end{equation}
where $S_{D4}$ represents the action of the D4-brane.  $S_{F1}$ is
the action of a stretched string from the boundary down to the
baryon vertex and $\tilde{S}_{F1}$ is the action of a radial
string hanging from the baryon vertex down to the horizon.  Recall
that $S_{D4}$ can be obtained from the Dirac-Born-Infeld
action\footnote{\begin{equation} S_{DBI}= \int dx^0 d\xi^p T_p ;
\quad T_p =
\left( e^{-\phi} (2\pi)^p {{\alpha}^\prime}^{(p+1)/2}\right)^{-1}\sqrt{-det(g)} \nonumber\\
\end{equation}}.  After some calculations, we obtain
\begin{equation}
S_{\text{D4}}=\frac{\tau N u_{c} \sqrt{f(u_{c})}}{6\pi \alpha^{'}
},\quad S_{\text{F1}}=\frac{\tau}{2\pi\alpha^{'} }\int_{0}^{L}
d\sigma\: \sqrt{{u^{\prime}}^2+f(u)\left( \frac{u}{R}\right) ^3},
\quad \tilde{S}_{\text{F1}}=\frac{\tau}{2\pi\alpha^{'}}
(u_{c}-u_{\text{T}}),\label{actions}
\end{equation}

\noindent where $\tau$ is the total time over which we evaluate
the action and $u_{c}$ is the position where the D4-brane vertex
is located.

The variation of the action with respect to $u$ gives the volume
term and the surface term.  The volume term leads to the usual
Euler-Lagrange equation for the classical configuration of
strings.  As an approximation, we assume the baryon vertex to be a
point~(not being distorted by the connecting strings) located at a
fixed value of $u=u_{c}$ as in Ref.~\cite{BISY}.  Under this
assumption, the surface terms provide additional {\it zero-force
condition} on the configuration,

\begin{equation}
\frac{N}{3}G_{0}(x) - k_{\text{h}} B +k_{\text{r}} =
0\label{no-force}
\end{equation}
\noindent where
\begin{equation}
 G_{0}(x) \equiv \frac{1+\frac{x^3}{2}}{\sqrt{1-x^3}} ,\quad
x\equiv \frac{u_{T}}{u_{c}} < 1, \;\text{and}\quad B \equiv
\frac{u_c^\prime}{\sqrt{{u_c^\prime}^2 + f(u_{c}) (
\frac{u_{c}}{R_{\text{D4}}}) ^3}}.\\ \label{B}
\end{equation}
\noindent Notice that these conditions occur at the location of
the vertex at $u=u_{c}$, at which there exists the balance between
the pull-up force (toward the direction of increasing $u$) due to
the tension of hanging strings and the pull-down force due to the
``weight"\footnote{This is not exactly the {\it weight} in the
usual sense since the direct gravitational force on Dbrane is
already balanced by the force from the RR-flux, but it is the
force originated from minimization of self-energy due to the brane
tension caused by the background metric and the gauge interaction.
This is very similar to the self-energy of a spring under gravity
where the spring potential energy changes with the tidal force
from gravity in the background.  The DBI action of the
D4$\thicksim u_{c}\sqrt{f(u_{c})}$ which is positive for
$u_{c}>u_{T}$ and becomes zero~(minimum) at $u_{c}=u_{T}$ and thus
it represents the ``weight" on D4 towards the horizon. } of
D4-brane plus the tension of radial strings.

Since  $ B \leq 1$, we obtain
\begin{equation}
 k_{\text{h}} \geq \frac{N}{3} G_{0}(x)+k_{\text{r}},\label{lowerbound}
\end{equation}

\noindent which expresses the lower bound of the number of hanging
strings. In other words, the number of hanging strings cannot be
less than this critical value, otherwise the no-force condition is
not satisfied.  The equality of \eqref{lowerbound} is held only
when all hanging strings are stretched straight, otherwise we
require more hanging strings to balance the pull-down force.
 Let us now consider each class of the multi-quark states.

In the case of {\bf $k$-baryon}, plugging the condition
\eqref{k-b_khkr} into \eqref{lowerbound}, we obtain
\begin{equation}
 k_{\text{h}}=k \geq \frac{N}{6}\left(G_0 (x)+3 \right).
\end{equation}
\noindent Apart from the lower bound, we also have the upper
bound, $k\leq N$, therefore $G_0 (x)$ cannot be larger than 3,
resulting in

\begin{equation}
 x\lesssim 0.922.
\end{equation}

\noindent Notice that this restriction on $x$ is a result from the
conditions of the force balance and conservation of string
charges. This shows that there is an upper-bound on the
temperature, over which the horizon is too near to the point
vertex that the pull-down force always overcomes the pull-up one.

In the case of {\bf{($N+\bar{k}$)-baryon}}, in the same way as the
preceding case, plugging the condition of charge conservation
\eqref{N+kbar_khkr} into \eqref{lowerbound}, we have the following
condition,
\begin{equation}
k_{\text{h}} =N+\bar{k}\geq \frac{N}{3} G_0 (x) +\bar{k}.
\nonumber
\end{equation}
\noindent  Unlike the case of $k$-baryon, the upper-bound of the
number of hanging strings does not exist.  However, we still
obtain the same condition $G_0 (x) \leq 3$, hence $x\lesssim
0.922$.

Finally, in the case of {\bf $j$-mesonance}, similarly,
Eqn.~\eqref{j-mkhkr} results in
\begin{equation}
 j\geq \frac{N}{6} \left( G_0 (x)+3\right).
\end{equation}
\noindent The lower-bound of the value of $j$ is $2N/3$ at zero
temperature ($x=0$) and it will be larger as the temperature
grows. Nevertheless, the upper-bound of the limit on $j$ does not
exist.

Finally, we would like to comment on the limits on the value of
$k, \bar{k}, j$ when the temperature is zero.  In terms of
$n\equiv 7-p$~(of the spacetime background generated by
D$p$-branes), the condition (\ref{lowerbound}) becomes
\begin{equation}
 k_{h}\geq \frac{N}{n} + k_{r}  \label{cond}
\end{equation}
which leads to
\begin{equation}
 \frac{k}{N}, \frac{j}{N}\geq \frac{n+1}{2n},
\end{equation}
and no conditions on $\bar{k}$.  This critical numbers are $5/8,
2/3$ for $n=4,3 $~(the AdS-Schwarzschild and Sakai-Sugimoto model)
respectively. It is an interesting coincidence that the critical
numbers are the same for both $k$-baryon and $j$-mesonance.  Even
though it appears from Eqn.~(\ref{cond}) that there should also be
a constraint on the $(N+\bar{k})$ configuration, it turns out that
there is none.

\section{Binding energy and the screening length}\label{energy}

In this section we will calculate the binding energies of the
$k$-baryon, $(N+\bar{k})$-baryon, and $j$-mesonance in the
deconfined phase.
 These binding energies are taken to be the differences between the
total energies of each configuration and the corresponding
energies of the free strings configuration which represents the
free quarks and/or antiquarks state.  The number of free strings
in the free quarks state is determined solely by the total number
of strings hanging from the boundary, $k_{h}$.

The total energy of each configuration is given by $E=S/\tau$ of
the corresponding action $S$ for each configuration.  The binding
energy for each hanging string is consequently,

\begin{equation}
E_{\text{F1}}=\frac{1}{2\pi } \int_{0}^{L} d\sigma\:
\sqrt{{u^{\prime}}^2 + \left(\frac{u}{R_{\text{D4}}}\right)^3
f(u)} -\frac{1}{2\pi}\int_{u_{T}}^{\infty}
du.\\
\end{equation}
Due to the no-force condition in the surface term, we impose
Eqn.~\eqref{no-force} and Eqn.~\eqref{B}, or

\begin{equation}\label{u0prime_n=3}
{u_c^\prime}^2 =\frac{f(u_c)  B^2}{1-B^2}\left(
\frac{u_c}{R_{\text{D4}}}\right)^3
\end{equation}

\noindent where the tension of each hanging string at $u_{c}$ is
constrained by
\begin{equation}
B=B(k_{\text{h}},k_{\text{r}},x)=\frac{N}{3 k_{\text{h}}} G_0
(x)+\frac{k_{\text{r}}}{k_{\text{h}}}.
\end{equation}

Since the Lagrangian $\mathcal{L}$ does not depend on $\sigma$
explicitly, the conserved Hamiltonian can be defined to be
\begin{equation}
\mathcal{H}\equiv\mathcal{L}-u^\prime \frac{\partial
\mathcal{L}}{\partial u^\prime}=\text{const},\\
\end{equation}
leading to
\begin{equation}
\frac{f(u_c)(\frac{u_c}{R_{\text{D4}}})^3}{\sqrt{{u_c^\prime}^2
+f(u_c)(\frac{u_c}{R_{\text{D4}}})^3}} =
\frac{f(u)(\frac{u}{R_{\text{D4}}})^3}{\sqrt{{u^\prime}^2
+f(u)(\frac{u}{R_{\text{D4}}})^3}}.
\end{equation}
Then substituting Eqn.~\eqref{u0prime_n=3} into this equation, we
obtain
\begin{equation}\label{uprime_n=3}
{u^\prime}^2 = \frac{f(u)^2(\frac{u}{R_{\text{D4}}})^6}{f(u_c)(
\frac{u_c}{R_{\text{D4}}})^3 (1-B^2)} -
f(u)\left(\frac{u}{R_{\text{D4}}}\right)^3.
\end{equation}
This gives the size (radius) of the baryon as seen on the gauge
theory side,
\begin{equation}
 L=\frac{R_{\text{D4}}^{3/2}}{u_c^{1/2}}\int_1^\infty dy\:
\sqrt{\frac{(1-x^3)(1-B^2)}{(y^3-x^3)(y^3-x^3-(1-x^3)(1-B^2))}}.\\
\end{equation}
Note that $u_c \approx \frac{R_{\text{D4}}^3}{L^2}$ at the leading
order.

Using Eqn.~\eqref{uprime_n=3} and let $y\equiv u/u_{c}$, the
regulated binding energy now becomes
\begin{equation}\label{EF1n=3}
E_{\text{F1}}=\frac{u_c}{2\pi } \Bigg\lbrace \int_1^\infty dy\:
\bigg\lbrack
\sqrt{\frac{y^3-x^3}{(y^3-x^3)-(1-x^3)(1-B^2)}}-1\bigg\rbrack-(1-x)
\Bigg\rbrace.
\end{equation}
Hence, we obtain the total energy of the configurations as
\begin{eqnarray}
E & = & \frac{N u_{T}}{2\pi
}\left(\frac{\sqrt{1-x^3}}{3x}+\left(\frac{k_\text{h}}{N}\right)\frac{\mathcal{E}}{x}
+\left(\frac{k_\text{r}}{N}\right)\frac{1-x}{x}\right)\label{Etot}
\\
& \thicksim & \frac{N^{2}}{L^{2}}
\end{eqnarray}
where $\mathcal{E}$ represents the terms within the brace of
\eqref{EF1n=3}.

\begin{figure}
\centering
\begingroup
  \makeatletter
  \providecommand\color[2][]{%
    \GenericError{(gnuplot) \space\space\space\@spaces}{%
      Package color not loaded in conjunction with
      terminal option `colourtext'%
    }{See the gnuplot documentation for explanation.%
    }{Either use 'blacktext' in gnuplot or load the package
      color.sty in LaTeX.}%
    \renewcommand\color[2][]{}%
  }%
  \providecommand\includegraphics[2][]{%
    \GenericError{(gnuplot) \space\space\space\@spaces}{%
      Package graphicx or graphics not loaded%
    }{See the gnuplot documentation for explanation.%
    }{The gnuplot epslatex terminal needs graphicx.sty or graphics.sty.}%
    \renewcommand\includegraphics[2][]{}%
  }%
  \providecommand\rotatebox[2]{#2}%
  \@ifundefined{ifGPcolor}{%
    \newif\ifGPcolor
    \GPcolorfalse
  }{}%
  \@ifundefined{ifGPblacktext}{%
    \newif\ifGPblacktext
    \GPblacktexttrue
  }{}%
  % define a \g@addto@macro without @ in the name:
  \let\gplgaddtomacro\g@addto@macro
  % define empty templates for all commands taking text:
  \gdef\gplbacktext{}%
  \gdef\gplfronttext{}%
  \makeatother
  \ifGPblacktext
    % no textcolor at all
    \def\colorrgb#1{}%
    \def\colorgray#1{}%
  \else
    % gray or color?
    \ifGPcolor
      \def\colorrgb#1{\color[rgb]{#1}}%
      \def\colorgray#1{\color[gray]{#1}}%
      \expandafter\def\csname LTw\endcsname{\color{white}}%
      \expandafter\def\csname LTb\endcsname{\color{black}}%
      \expandafter\def\csname LTa\endcsname{\color{black}}%
      \expandafter\def\csname LT0\endcsname{\color[rgb]{1,0,0}}%
      \expandafter\def\csname LT1\endcsname{\color[rgb]{0,1,0}}%
      \expandafter\def\csname LT2\endcsname{\color[rgb]{0,0,1}}%
      \expandafter\def\csname LT3\endcsname{\color[rgb]{1,0,1}}%
      \expandafter\def\csname LT4\endcsname{\color[rgb]{0,1,1}}%
      \expandafter\def\csname LT5\endcsname{\color[rgb]{1,1,0}}%
      \expandafter\def\csname LT6\endcsname{\color[rgb]{0,0,0}}%
      \expandafter\def\csname LT7\endcsname{\color[rgb]{1,0.3,0}}%
      \expandafter\def\csname LT8\endcsname{\color[rgb]{0.5,0.5,0.5}}%
    \else
      % gray
      \def\colorrgb#1{\color{black}}%
      \def\colorgray#1{\color[gray]{#1}}%
      \expandafter\def\csname LTw\endcsname{\color{white}}%
      \expandafter\def\csname LTb\endcsname{\color{black}}%
      \expandafter\def\csname LTa\endcsname{\color{black}}%
      \expandafter\def\csname LT0\endcsname{\color{black}}%
      \expandafter\def\csname LT1\endcsname{\color{black}}%
      \expandafter\def\csname LT2\endcsname{\color{black}}%
      \expandafter\def\csname LT3\endcsname{\color{black}}%
      \expandafter\def\csname LT4\endcsname{\color{black}}%
      \expandafter\def\csname LT5\endcsname{\color{black}}%
      \expandafter\def\csname LT6\endcsname{\color{black}}%
      \expandafter\def\csname LT7\endcsname{\color{black}}%
      \expandafter\def\csname LT8\endcsname{\color{black}}%
    \fi
  \fi
  \setlength{\unitlength}{0.0500bp}%
  \begin{picture}(7200.00,5040.00)%
    \gplgaddtomacro\gplbacktext{%
      \csname LTb\endcsname%
      \put(1188,717){\makebox(0,0)[r]{\strut{}$-0.25$}}%
      \put(1188,1332){\makebox(0,0)[r]{\strut{}$-0.20$}}%
      \put(1188,1947){\makebox(0,0)[r]{\strut{}$-0.15$}}%
      \put(1188,2562){\makebox(0,0)[r]{\strut{}$-0.10$}}%
      \put(1188,3177){\makebox(0,0)[r]{\strut{}$-0.05$}}%
      \put(1188,3792){\makebox(0,0)[r]{\strut{}$ 0.00$}}%
      \put(1188,4407){\makebox(0,0)[r]{\strut{}$ 0.05$}}%
      \put(1320,374){\makebox(0,0){\strut{}$ 0.00$}}%
      \put(2517,374){\makebox(0,0){\strut{}$ 0.10$}}%
      \put(3714,374){\makebox(0,0){\strut{}$ 0.20$}}%
      \put(4911,374){\makebox(0,0){\strut{}$ 0.30$}}%
      \put(6108,374){\makebox(0,0){\strut{}$ 0.40$}}%
      \csname LTb\endcsname%
      \put(220,2685){\makebox(0,0){$E/N$}}%
      \put(4073,110){\makebox(0,0){$L$}}%
    }%
    \gplgaddtomacro\gplfronttext{%
      \csname LTb\endcsname%
      \put(5732,1591){\makebox(0,0)[r]{\strut{}$N$}}%
      \csname LTb\endcsname%
      \put(5732,1371){\makebox(0,0)[r]{\strut{}$k$}}%
      \csname LTb\endcsname%
      \put(5732,1151){\makebox(0,0)[r]{\strut{}$N+\bar{k}_1$}}%
      \csname LTb\endcsname%
      \put(5732,931){\makebox(0,0)[r]{\strut{}$N+\bar{k}_2$}}%
    }%
    \gplbacktext
    \put(0,0){\includegraphics{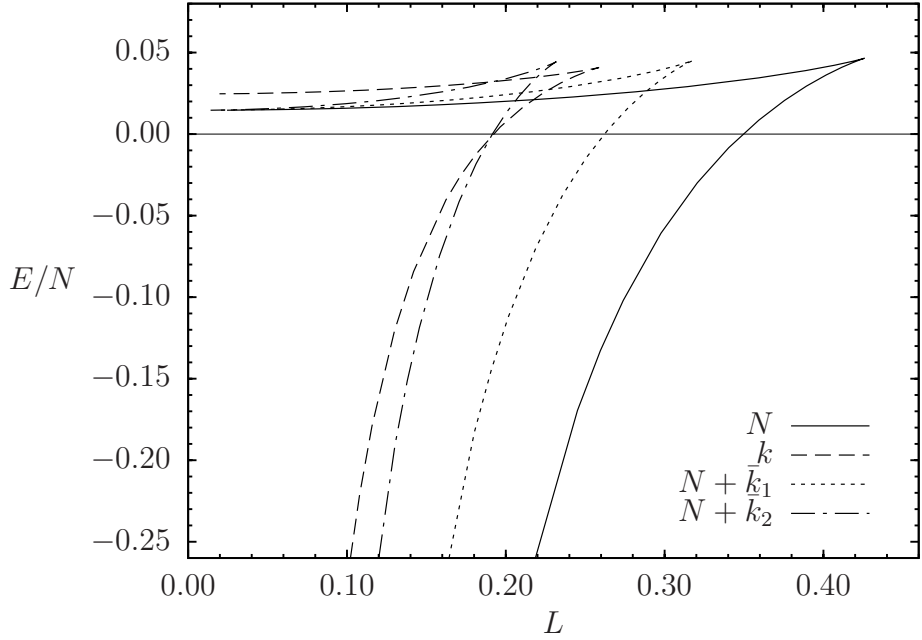}}%
    \gplfronttext
  \end{picture}%
\endgroup
\caption{Comparison of the potential per $N$ between $N$-baryon,
$k$-baryon, and $(N+\bar{k})$-baryon for $k/N=0.8,
\bar{k_{1}}/N=2/3, \bar{k_{2}}/N=2$ at temperature $T= 0.25$.}
\label{ceb}
\end{figure}

\begin{figure}
\centering
\begingroup
  \makeatletter
  \providecommand\color[2][]{%
    \GenericError{(gnuplot) \space\space\space\@spaces}{%
      Package color not loaded in conjunction with
      terminal option `colourtext'%
    }{See the gnuplot documentation for explanation.%
    }{Either use 'blacktext' in gnuplot or load the package
      color.sty in LaTeX.}%
    \renewcommand\color[2][]{}%
  }%
  \providecommand\includegraphics[2][]{%
    \GenericError{(gnuplot) \space\space\space\@spaces}{%
      Package graphicx or graphics not loaded%
    }{See the gnuplot documentation for explanation.%
    }{The gnuplot epslatex terminal needs graphicx.sty or graphics.sty.}%
    \renewcommand\includegraphics[2][]{}%
  }%
  \providecommand\rotatebox[2]{#2}%
  \@ifundefined{ifGPcolor}{%
    \newif\ifGPcolor
    \GPcolorfalse
  }{}%
  \@ifundefined{ifGPblacktext}{%
    \newif\ifGPblacktext
    \GPblacktexttrue
  }{}%
  % define a \g@addto@macro without @ in the name:
  \let\gplgaddtomacro\g@addto@macro
  % define empty templates for all commands taking text:
  \gdef\gplbacktext{}%
  \gdef\gplfronttext{}%
  \makeatother
  \ifGPblacktext
    % no textcolor at all
    \def\colorrgb#1{}%
    \def\colorgray#1{}%
  \else
    % gray or color?
    \ifGPcolor
      \def\colorrgb#1{\color[rgb]{#1}}%
      \def\colorgray#1{\color[gray]{#1}}%
      \expandafter\def\csname LTw\endcsname{\color{white}}%
      \expandafter\def\csname LTb\endcsname{\color{black}}%
      \expandafter\def\csname LTa\endcsname{\color{black}}%
      \expandafter\def\csname LT0\endcsname{\color[rgb]{1,0,0}}%
      \expandafter\def\csname LT1\endcsname{\color[rgb]{0,1,0}}%
      \expandafter\def\csname LT2\endcsname{\color[rgb]{0,0,1}}%
      \expandafter\def\csname LT3\endcsname{\color[rgb]{1,0,1}}%
      \expandafter\def\csname LT4\endcsname{\color[rgb]{0,1,1}}%
      \expandafter\def\csname LT5\endcsname{\color[rgb]{1,1,0}}%
      \expandafter\def\csname LT6\endcsname{\color[rgb]{0,0,0}}%
      \expandafter\def\csname LT7\endcsname{\color[rgb]{1,0.3,0}}%
      \expandafter\def\csname LT8\endcsname{\color[rgb]{0.5,0.5,0.5}}%
    \else
      % gray
      \def\colorrgb#1{\color{black}}%
      \def\colorgray#1{\color[gray]{#1}}%
      \expandafter\def\csname LTw\endcsname{\color{white}}%
      \expandafter\def\csname LTb\endcsname{\color{black}}%
      \expandafter\def\csname LTa\endcsname{\color{black}}%
      \expandafter\def\csname LT0\endcsname{\color{black}}%
      \expandafter\def\csname LT1\endcsname{\color{black}}%
      \expandafter\def\csname LT2\endcsname{\color{black}}%
      \expandafter\def\csname LT3\endcsname{\color{black}}%
      \expandafter\def\csname LT4\endcsname{\color{black}}%
      \expandafter\def\csname LT5\endcsname{\color{black}}%
      \expandafter\def\csname LT6\endcsname{\color{black}}%
      \expandafter\def\csname LT7\endcsname{\color{black}}%
      \expandafter\def\csname LT8\endcsname{\color{black}}%
    \fi
  \fi
  \setlength{\unitlength}{0.0500bp}%
  \begin{picture}(7200.00,5040.00)%
    \gplgaddtomacro\gplbacktext{%
      \csname LTb\endcsname%
      \put(1166,803){\makebox(0,0)[r]{\strut{}$ -1.5$}}%
      \put(1166,1849){\makebox(0,0)[r]{\strut{}$ -1.0$}}%
      \put(1166,2894){\makebox(0,0)[r]{\strut{}$ -0.5$}}%
      \put(1166,3940){\makebox(0,0)[r]{\strut{}$  0.0$}}%
      \put(1298,374){\makebox(0,0){\strut{}$ 0.00$}}%
      \put(2066,374){\makebox(0,0){\strut{}$ 0.10$}}%
      \put(2834,374){\makebox(0,0){\strut{}$ 0.20$}}%
      \put(3601,374){\makebox(0,0){\strut{}$ 0.30$}}%
      \put(4369,374){\makebox(0,0){\strut{}$ 0.40$}}%
      \put(5137,374){\makebox(0,0){\strut{}$ 0.50$}}%
      \put(5905,374){\makebox(0,0){\strut{}$ 0.60$}}%
      \put(6672,374){\makebox(0,0){\strut{}$ 0.70$}}%
      \csname LTb\endcsname%
      \put(396,2685){\makebox(0,0){$E/N$}}%
      \put(4062,110){\makebox(0,0){$L$}}%
    }%
    \gplgaddtomacro\gplfronttext{%
      \csname LTb\endcsname%
      \put(5817,1530){\makebox(0,0)[r]{\strut{}$j_1$ mesons}}%
      \csname LTb\endcsname%
      \put(5817,1310){\makebox(0,0)[r]{\strut{}$j_1$-mesonance}}%
      \csname LTb\endcsname%
      \put(5817,1090){\makebox(0,0)[r]{\strut{}$j_2$ mesons}}%
      \csname LTb\endcsname%
      \put(5817,870){\makebox(0,0)[r]{\strut{}$j_2$-mesonance}}%
    }%
    \gplbacktext
    \put(0,0){\includegraphics{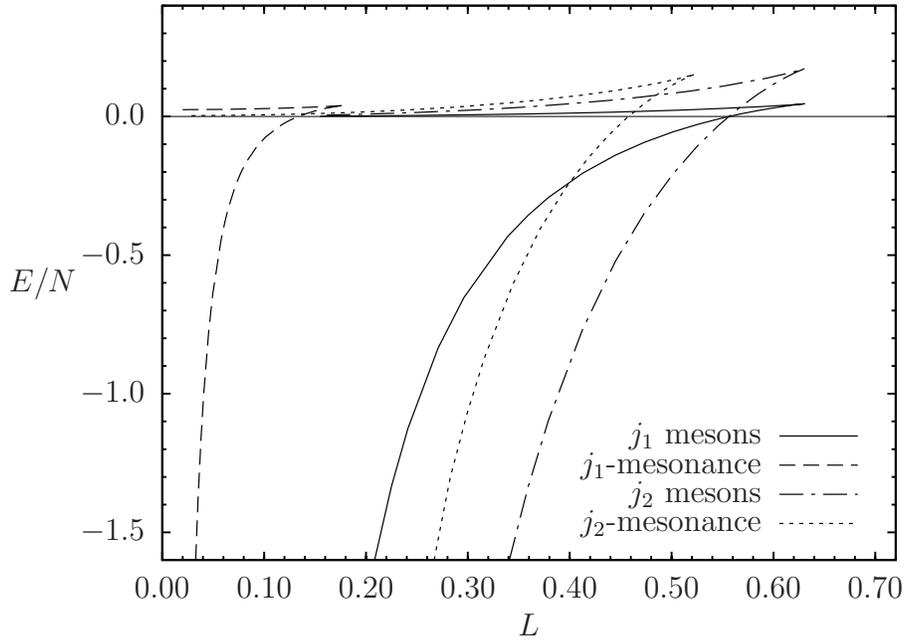}}%
    \gplfronttext
  \end{picture}%
\endgroup
\caption{Comparison of the potential per $N$ between $j$-mesonance
and $j$ mesons for $j_{1}/N=0.8, j_{2}/N=3$ at temperature $T=
0.25$.} \label{cem}
\end{figure}

To obtain the relations between the total energy of the
configurations $E(x)$ and $L(x)$, we eliminate the parameter
$x=u_{T}/u_{c}$.  By numerical calculations, the results are shown
in Fig.~\ref{ceb},\ref{cem}.  The binding energy of $N$-baryon is
the deepest, suggesting that it is the most tightly bound state.
For $(N+\bar{k})$-baryon, increasing $\bar{k}$ makes the binding
energy smaller and the bound state is less tightly bound.  The
case of $j$-mesonance is quite similar.  Generically, a
$j$-mesonance has shallower binding potential than the total
energy of $j$ mesons.  However, as $j$ grows, the difference gets
smaller and smaller.
%\input{baryons3.tex}
% GNUPLOT: LaTeX picture with Postscript

%\input{j_m3.tex}
% GNUPLOT: LaTeX picture with Postscript

The screening radius or screening length of exotic multi-quark
state is defined to be the value of radius $L^{*}$ at which the
binding energy becomes zero from negative values at smaller
distances. This screening radius is therefore one-half of the
usual definition of screening length in the discussion of mesonic
state where it is defined as the zero-potential distance between
quark and antiquark.

\begin{figure}
\centering
\begingroup
  \makeatletter
  \providecommand\color[2][]{%
    \GenericError{(gnuplot) \space\space\space\@spaces}{%
      Package color not loaded in conjunction with
      terminal option `colourtext'%
    }{See the gnuplot documentation for explanation.%
    }{Either use 'blacktext' in gnuplot or load the package
      color.sty in LaTeX.}%
    \renewcommand\color[2][]{}%
  }%
  \providecommand\includegraphics[2][]{%
    \GenericError{(gnuplot) \space\space\space\@spaces}{%
      Package graphicx or graphics not loaded%
    }{See the gnuplot documentation for explanation.%
    }{The gnuplot epslatex terminal needs graphicx.sty or graphics.sty.}%
    \renewcommand\includegraphics[2][]{}%
  }%
  \providecommand\rotatebox[2]{#2}%
  \@ifundefined{ifGPcolor}{%
    \newif\ifGPcolor
    \GPcolorfalse
  }{}%
  \@ifundefined{ifGPblacktext}{%
    \newif\ifGPblacktext
    \GPblacktexttrue
  }{}%
  % define a \g@addto@macro without @ in the name:
  \let\gplgaddtomacro\g@addto@macro
  % define empty templates for all commands taking text:
  \gdef\gplbacktext{}%
  \gdef\gplfronttext{}%
  \makeatother
  \ifGPblacktext
    % no textcolor at all
    \def\colorrgb#1{}%
    \def\colorgray#1{}%
  \else
    % gray or color?
    \ifGPcolor
      \def\colorrgb#1{\color[rgb]{#1}}%
      \def\colorgray#1{\color[gray]{#1}}%
      \expandafter\def\csname LTw\endcsname{\color{white}}%
      \expandafter\def\csname LTb\endcsname{\color{black}}%
      \expandafter\def\csname LTa\endcsname{\color{black}}%
      \expandafter\def\csname LT0\endcsname{\color[rgb]{1,0,0}}%
      \expandafter\def\csname LT1\endcsname{\color[rgb]{0,1,0}}%
      \expandafter\def\csname LT2\endcsname{\color[rgb]{0,0,1}}%
      \expandafter\def\csname LT3\endcsname{\color[rgb]{1,0,1}}%
      \expandafter\def\csname LT4\endcsname{\color[rgb]{0,1,1}}%
      \expandafter\def\csname LT5\endcsname{\color[rgb]{1,1,0}}%
      \expandafter\def\csname LT6\endcsname{\color[rgb]{0,0,0}}%
      \expandafter\def\csname LT7\endcsname{\color[rgb]{1,0.3,0}}%
      \expandafter\def\csname LT8\endcsname{\color[rgb]{0.5,0.5,0.5}}%
    \else
      % gray
      \def\colorrgb#1{\color{black}}%
      \def\colorgray#1{\color[gray]{#1}}%
      \expandafter\def\csname LTw\endcsname{\color{white}}%
      \expandafter\def\csname LTb\endcsname{\color{black}}%
      \expandafter\def\csname LTa\endcsname{\color{black}}%
      \expandafter\def\csname LT0\endcsname{\color{black}}%
      \expandafter\def\csname LT1\endcsname{\color{black}}%
      \expandafter\def\csname LT2\endcsname{\color{black}}%
      \expandafter\def\csname LT3\endcsname{\color{black}}%
      \expandafter\def\csname LT4\endcsname{\color{black}}%
      \expandafter\def\csname LT5\endcsname{\color{black}}%
      \expandafter\def\csname LT6\endcsname{\color{black}}%
      \expandafter\def\csname LT7\endcsname{\color{black}}%
      \expandafter\def\csname LT8\endcsname{\color{black}}%
    \fi
  \fi
  \setlength{\unitlength}{0.0500bp}%
  \begin{picture}(7200.00,5040.00)%
    \gplgaddtomacro\gplbacktext{%
      \csname LTb\endcsname%
      \put(1166,594){\makebox(0,0)[r]{\strut{}$  0.0$}}%
      \put(1166,1291){\makebox(0,0)[r]{\strut{}$  0.1$}}%
      \put(1166,1988){\makebox(0,0)[r]{\strut{}$  0.2$}}%
      \put(1166,2685){\makebox(0,0)[r]{\strut{}$  0.3$}}%
      \put(1166,3382){\makebox(0,0)[r]{\strut{}$  0.4$}}%
      \put(1166,4079){\makebox(0,0)[r]{\strut{}$  0.5$}}%
      \put(1166,4776){\makebox(0,0)[r]{\strut{}$  0.6$}}%
      \put(1298,374){\makebox(0,0){\strut{}$ 0.65$}}%
      \put(2088,374){\makebox(0,0){\strut{}$ 0.70$}}%
      \put(2877,374){\makebox(0,0){\strut{}$ 0.75$}}%
      \put(3667,374){\makebox(0,0){\strut{}$ 0.80$}}%
      \put(4457,374){\makebox(0,0){\strut{}$ 0.85$}}%
      \put(5247,374){\makebox(0,0){\strut{}$ 0.90$}}%
      \put(6036,374){\makebox(0,0){\strut{}$ 0.95$}}%
      \put(6826,374){\makebox(0,0){\strut{}$ 1.00$}}%
      \csname LTb\endcsname%
      \put(528,2685){\makebox(0,0){$L^{*}$}}%
      \put(4062,110){\makebox(0,0){$k/N$}}%
    }%
    \gplgaddtomacro\gplfronttext{%
      \csname LTb\endcsname%
      \put(5813,1669){\makebox(0,0)[r]{\strut{}$T=0.15$}}%
      \csname LTb\endcsname%
      \put(5813,1449){\makebox(0,0)[r]{\strut{}$T=0.20$}}%
      \csname LTb\endcsname%
      \put(5813,1229){\makebox(0,0)[r]{\strut{}$T=0.25$}}%
      \csname LTb\endcsname%
      \put(5813,1009){\makebox(0,0)[r]{\strut{}$T=0.30$}}%
      \csname LTb\endcsname%
      \put(5813,789){\makebox(0,0)[r]{\strut{}$T=0.35$}}%
    }%
    \gplbacktext
    \put(0,0){\includegraphics{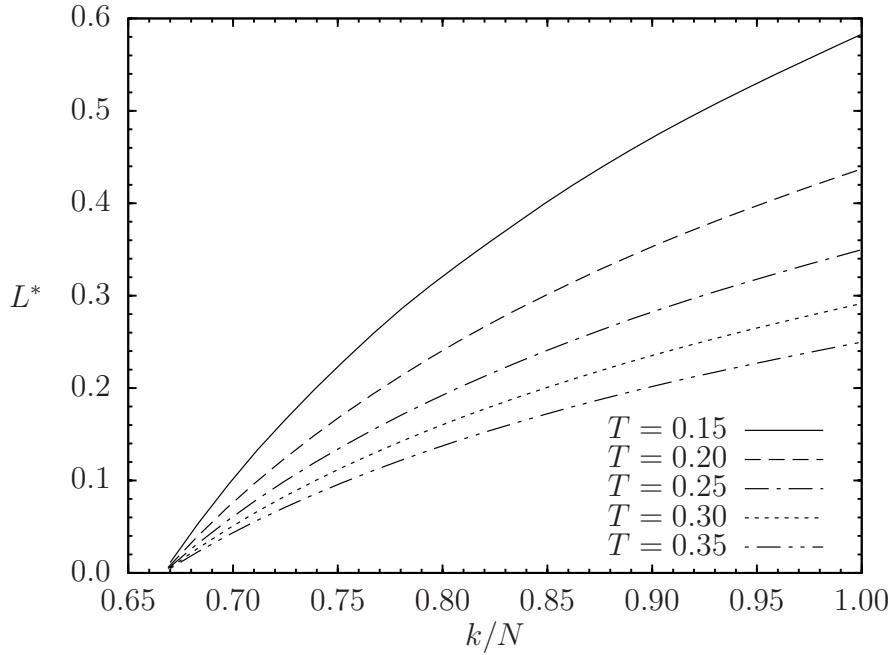}}%
    \gplfronttext
  \end{picture}%
\endgroup
\caption{Screening length with respect to $k$ for the temperatures
in $0.15-0.35$ range.} \label{slk}
\end{figure}

\begin{figure}
\centering
\begingroup
  \makeatletter
  \providecommand\color[2][]{%
    \GenericError{(gnuplot) \space\space\space\@spaces}{%
      Package color not loaded in conjunction with
      terminal option `colourtext'%
    }{See the gnuplot documentation for explanation.%
    }{Either use 'blacktext' in gnuplot or load the package
      color.sty in LaTeX.}%
    \renewcommand\color[2][]{}%
  }%
  \providecommand\includegraphics[2][]{%
    \GenericError{(gnuplot) \space\space\space\@spaces}{%
      Package graphicx or graphics not loaded%
    }{See the gnuplot documentation for explanation.%
    }{The gnuplot epslatex terminal needs graphicx.sty or graphics.sty.}%
    \renewcommand\includegraphics[2][]{}%
  }%
  \providecommand\rotatebox[2]{#2}%
  \@ifundefined{ifGPcolor}{%
    \newif\ifGPcolor
    \GPcolorfalse
  }{}%
  \@ifundefined{ifGPblacktext}{%
    \newif\ifGPblacktext
    \GPblacktexttrue
  }{}%
  % define a \g@addto@macro without @ in the name:
  \let\gplgaddtomacro\g@addto@macro
  % define empty templates for all commands taking text:
  \gdef\gplbacktext{}%
  \gdef\gplfronttext{}%
  \makeatother
  \ifGPblacktext
    % no textcolor at all
    \def\colorrgb#1{}%
    \def\colorgray#1{}%
  \else
    % gray or color?
    \ifGPcolor
      \def\colorrgb#1{\color[rgb]{#1}}%
      \def\colorgray#1{\color[gray]{#1}}%
      \expandafter\def\csname LTw\endcsname{\color{white}}%
      \expandafter\def\csname LTb\endcsname{\color{black}}%
      \expandafter\def\csname LTa\endcsname{\color{black}}%
      \expandafter\def\csname LT0\endcsname{\color[rgb]{1,0,0}}%
      \expandafter\def\csname LT1\endcsname{\color[rgb]{0,1,0}}%
      \expandafter\def\csname LT2\endcsname{\color[rgb]{0,0,1}}%
      \expandafter\def\csname LT3\endcsname{\color[rgb]{1,0,1}}%
      \expandafter\def\csname LT4\endcsname{\color[rgb]{0,1,1}}%
      \expandafter\def\csname LT5\endcsname{\color[rgb]{1,1,0}}%
      \expandafter\def\csname LT6\endcsname{\color[rgb]{0,0,0}}%
      \expandafter\def\csname LT7\endcsname{\color[rgb]{1,0.3,0}}%
      \expandafter\def\csname LT8\endcsname{\color[rgb]{0.5,0.5,0.5}}%
    \else
      % gray
      \def\colorrgb#1{\color{black}}%
      \def\colorgray#1{\color[gray]{#1}}%
      \expandafter\def\csname LTw\endcsname{\color{white}}%
      \expandafter\def\csname LTb\endcsname{\color{black}}%
      \expandafter\def\csname LTa\endcsname{\color{black}}%
      \expandafter\def\csname LT0\endcsname{\color{black}}%
      \expandafter\def\csname LT1\endcsname{\color{black}}%
      \expandafter\def\csname LT2\endcsname{\color{black}}%
      \expandafter\def\csname LT3\endcsname{\color{black}}%
      \expandafter\def\csname LT4\endcsname{\color{black}}%
      \expandafter\def\csname LT5\endcsname{\color{black}}%
      \expandafter\def\csname LT6\endcsname{\color{black}}%
      \expandafter\def\csname LT7\endcsname{\color{black}}%
      \expandafter\def\csname LT8\endcsname{\color{black}}%
    \fi
  \fi
  \setlength{\unitlength}{0.0500bp}%
  \begin{picture}(7200.00,5040.00)%
    \gplgaddtomacro\gplbacktext{%
      \csname LTb\endcsname%
      \put(1166,594){\makebox(0,0)[r]{\strut{}$  0.0$}}%
      \put(1166,1291){\makebox(0,0)[r]{\strut{}$  0.1$}}%
      \put(1166,1988){\makebox(0,0)[r]{\strut{}$  0.2$}}%
      \put(1166,2685){\makebox(0,0)[r]{\strut{}$  0.3$}}%
      \put(1166,3382){\makebox(0,0)[r]{\strut{}$  0.4$}}%
      \put(1166,4079){\makebox(0,0)[r]{\strut{}$  0.5$}}%
      \put(1166,4776){\makebox(0,0)[r]{\strut{}$  0.6$}}%
      \put(1298,374){\makebox(0,0){\strut{}$    0$}}%
      \put(1989,374){\makebox(0,0){\strut{}$    5$}}%
      \put(2680,374){\makebox(0,0){\strut{}$   10$}}%
      \put(3371,374){\makebox(0,0){\strut{}$   15$}}%
      \put(4062,374){\makebox(0,0){\strut{}$   20$}}%
      \put(4753,374){\makebox(0,0){\strut{}$   25$}}%
      \put(5444,374){\makebox(0,0){\strut{}$   30$}}%
      \put(6135,374){\makebox(0,0){\strut{}$   35$}}%
      \put(6826,374){\makebox(0,0){\strut{}$   40$}}%
      \csname LTb\endcsname%
      \put(528,2685){\makebox(0,0){$L^{*}$}}%
      \put(4062,110){\makebox(0,0){$\bar{k}/N$}}%
    }%
    \gplgaddtomacro\gplfronttext{%
      \csname LTb\endcsname%
      \put(5902,4527){\makebox(0,0)[r]{\strut{}$T=0.15$}}%
      \csname LTb\endcsname%
      \put(5902,4307){\makebox(0,0)[r]{\strut{}$T=0.20$}}%
      \csname LTb\endcsname%
      \put(5902,4087){\makebox(0,0)[r]{\strut{}$T=0.25$}}%
      \csname LTb\endcsname%
      \put(5902,3867){\makebox(0,0)[r]{\strut{}$T=0.30$}}%
      \csname LTb\endcsname%
      \put(5902,3647){\makebox(0,0)[r]{\strut{}$T=0.35$}}%
    }%
    \gplbacktext
    \put(0,0){\includegraphics{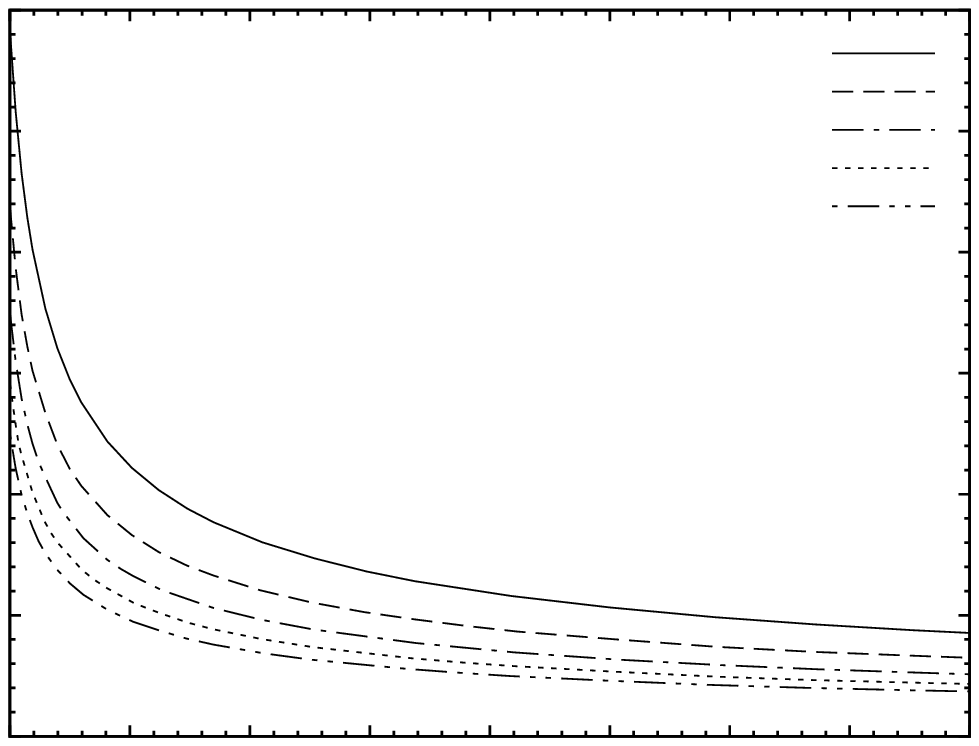}}%
    \gplfronttext
  \end{picture}%
\endgroup
\caption{Screening length with respect to $\bar{k}$ for the
temperatures in $0.15-0.35$ range.} \label{slkb}
\end{figure}

\begin{figure}
\centering
\begingroup
  \makeatletter
  \providecommand\color[2][]{%
    \GenericError{(gnuplot) \space\space\space\@spaces}{%
      Package color not loaded in conjunction with
      terminal option `colourtext'%
    }{See the gnuplot documentation for explanation.%
    }{Either use 'blacktext' in gnuplot or load the package
      color.sty in LaTeX.}%
    \renewcommand\color[2][]{}%
  }%
  \providecommand\includegraphics[2][]{%
    \GenericError{(gnuplot) \space\space\space\@spaces}{%
      Package graphicx or graphics not loaded%
    }{See the gnuplot documentation for explanation.%
    }{The gnuplot epslatex terminal needs graphicx.sty or graphics.sty.}%
    \renewcommand\includegraphics[2][]{}%
  }%
  \providecommand\rotatebox[2]{#2}%
  \@ifundefined{ifGPcolor}{%
    \newif\ifGPcolor
    \GPcolorfalse
  }{}%
  \@ifundefined{ifGPblacktext}{%
    \newif\ifGPblacktext
    \GPblacktexttrue
  }{}%
  % define a \g@addto@macro without @ in the name:
  \let\gplgaddtomacro\g@addto@macro
  % define empty templates for all commands taking text:
  \gdef\gplbacktext{}%
  \gdef\gplfronttext{}%
  \makeatother
  \ifGPblacktext
    % no textcolor at all
    \def\colorrgb#1{}%
    \def\colorgray#1{}%
  \else
    % gray or color?
    \ifGPcolor
      \def\colorrgb#1{\color[rgb]{#1}}%
      \def\colorgray#1{\color[gray]{#1}}%
      \expandafter\def\csname LTw\endcsname{\color{white}}%
      \expandafter\def\csname LTb\endcsname{\color{black}}%
      \expandafter\def\csname LTa\endcsname{\color{black}}%
      \expandafter\def\csname LT0\endcsname{\color[rgb]{1,0,0}}%
      \expandafter\def\csname LT1\endcsname{\color[rgb]{0,1,0}}%
      \expandafter\def\csname LT2\endcsname{\color[rgb]{0,0,1}}%
      \expandafter\def\csname LT3\endcsname{\color[rgb]{1,0,1}}%
      \expandafter\def\csname LT4\endcsname{\color[rgb]{0,1,1}}%
      \expandafter\def\csname LT5\endcsname{\color[rgb]{1,1,0}}%
      \expandafter\def\csname LT6\endcsname{\color[rgb]{0,0,0}}%
      \expandafter\def\csname LT7\endcsname{\color[rgb]{1,0.3,0}}%
      \expandafter\def\csname LT8\endcsname{\color[rgb]{0.5,0.5,0.5}}%
    \else
      % gray
      \def\colorrgb#1{\color{black}}%
      \def\colorgray#1{\color[gray]{#1}}%
      \expandafter\def\csname LTw\endcsname{\color{white}}%
      \expandafter\def\csname LTb\endcsname{\color{black}}%
      \expandafter\def\csname LTa\endcsname{\color{black}}%
      \expandafter\def\csname LT0\endcsname{\color{black}}%
      \expandafter\def\csname LT1\endcsname{\color{black}}%
      \expandafter\def\csname LT2\endcsname{\color{black}}%
      \expandafter\def\csname LT3\endcsname{\color{black}}%
      \expandafter\def\csname LT4\endcsname{\color{black}}%
      \expandafter\def\csname LT5\endcsname{\color{black}}%
      \expandafter\def\csname LT6\endcsname{\color{black}}%
      \expandafter\def\csname LT7\endcsname{\color{black}}%
      \expandafter\def\csname LT8\endcsname{\color{black}}%
    \fi
  \fi
  \setlength{\unitlength}{0.0500bp}%
  \begin{picture}(7200.00,5040.00)%
    \gplgaddtomacro\gplbacktext{%
      \csname LTb\endcsname%
      \put(1166,594){\makebox(0,0)[r]{\strut{}$  0.0$}}%
      \put(1166,1484){\makebox(0,0)[r]{\strut{}$  0.2$}}%
      \put(1166,2374){\makebox(0,0)[r]{\strut{}$  0.4$}}%
      \put(1166,3263){\makebox(0,0)[r]{\strut{}$  0.6$}}%
      \put(1166,4153){\makebox(0,0)[r]{\strut{}$  0.8$}}%
      \put(1298,374){\makebox(0,0){\strut{}$    0$}}%
      \put(2680,374){\makebox(0,0){\strut{}$   10$}}%
      \put(4062,374){\makebox(0,0){\strut{}$   20$}}%
      \put(5444,374){\makebox(0,0){\strut{}$   30$}}%
      \put(6826,374){\makebox(0,0){\strut{}$   40$}}%
      \csname LTb\endcsname%
      \put(528,2685){\makebox(0,0){$L^{*}$}}%
      \put(4062,110){\makebox(0,0){$j/N$}}%
    }%
    \gplgaddtomacro\gplfronttext{%
      \csname LTb\endcsname%
      \put(5833,1730){\makebox(0,0)[r]{\strut{}$T=0.15$}}%
      \csname LTb\endcsname%
      \put(5833,1510){\makebox(0,0)[r]{\strut{}$T=0.20$}}%
      \csname LTb\endcsname%
      \put(5833,1290){\makebox(0,0)[r]{\strut{}$T=0.25$}}%
      \csname LTb\endcsname%
      \put(5833,1070){\makebox(0,0)[r]{\strut{}$T=0.30$}}%
      \csname LTb\endcsname%
      \put(5833,850){\makebox(0,0)[r]{\strut{}$T=0.35$}}%
    }%
    \gplbacktext
    \put(0,0){\includegraphics{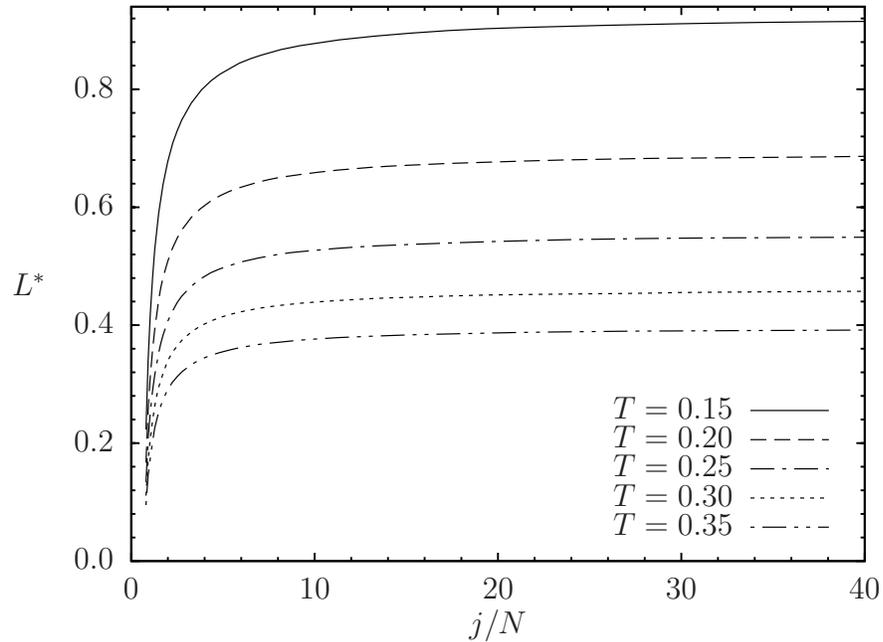}}%
    \gplfronttext
  \end{picture}%
\endgroup
\caption{Screening length with respect to $j$ for the temperatures
in $0.15-0.35$ range.} \label{slj}
\end{figure}

Numerical results suggest that the screening length of baryons and
mesonance decrease as the temperature increases, i.e. $L^{*} \sim
1/T$ for a fixed value of $k,\bar{k},j$ as is shown in
Fig.~\ref{slk}-\ref{slj}. This is the generic form for the
screening length in both the AdS-Schwarzschild and Sakai-Sugimoto
models because it is the quantity which does not depend on the 't
Hooft coupling at the leading order~\cite{bl}.
 It is also an increasing function of $k$ and $j$. Interestingly,
$(N+\bar{k})$-baryon has the opposite tendency with the screening
length decreases as $\bar{k}$ grows. On the other hand, the
screening length of $j$-mesonance has a saturation value
$L^{*}_{j-mesonance}\to L^{*}_{meson}$ as $j \to \infty$.

\section{Dependence on the free quark mass} \label{mass dependence}

In this section, we will study dependence of the binding potential
on the position of the probe branes.  This is useful when position
of the probe branes are at finite distance from the black hole
horizon and the corresponding quarks have finite mass.  For
example, the probe branes are D8 and $\overline{\text{D8}}$
flavour branes in the Sakai-Sugimoto model.

The calculation of binding energy as a function of the radius $L$
of the multi-quark states in the previous sections can be
generalized to the case where the background metric is generated
by a stack of D$p$-branes as the following.  Start with the energy
of a hanging fundamental string with $n=7-p$,

\begin{equation}\label{EF1rjlarge}
E_{F1}=\frac{u_c}{2\pi} \Bigg\lbrace \int_1^\infty dy\:
\bigg\lbrack
\sqrt{\frac{y^n -x^n}{(y^n-x^n)-(1-x^n)(1-{A(n)}^2)}}-1\bigg\rbrack-(1-x) \Bigg\rbrace\\
\end{equation}
and the radius,
\begin{equation}\label{Lrjlarge}
 L=\frac{R^{n/2}}{{u_c}^{(n-2)/2}}\int_1^\infty dy\:
\sqrt{\frac{(1-x^n)(1-A(n)^2)}{(y^n-x^n)(y^n-x^n-(1-x^n)(1-A(n)^2))}}.\\
\end{equation}
The total regulated binding energy of the configuration then
becomes
\begin{equation}\label{Etotrj}
E_{tot} = \frac{N
u_h}{2\pi}\Bigg\lbrace\frac{\sqrt{1-x^n}}{nx}+\left(
\frac{k_\text{h}}{N}\right)\frac{\mathcal{E}}{x}
+\left(\frac{k_\text{r}}{N}\right)\frac{1-x}{x}\Bigg\rbrace
\end{equation}

where

\begin{equation}
\mathcal{E}= \int_1^\infty dy\: \bigg[
\sqrt{\frac{y^n-x^n}{(y^n-x^n)-(1-x^n)(1-{A(n)}^2)}}-1\bigg]
-(1-x),\label{mathcalE}
\end{equation}

 and

\begin{equation}
A(n)= \frac{u_c^\prime}{\sqrt{{u_c^\prime}^2 + f(u_{c}) (
\frac{u_{c}}{R_{\text{D}p}}) ^n}} = \frac{N}{n k_{\text{h}}}\left(
\frac{1+
\frac{n-2}{2}x^{n}}{\sqrt{1-x^{n}}}\right)+\frac{k_{\text{r}}}{k_{\text{h}}}.\label{An}
\end{equation}
The parameter $x$ is again given by
\begin{equation}
x=\frac{{u_T}(n)}{u_c}, \quad u_T (n=3,4) =\frac{16}{9}{\pi}^2 R^3
T^2, \pi R^2 T.
\end{equation}
Note that the case $n=3$ and $n=4$ corresponds to the case of
Sakai-Sugimoto and AdS-Schwarzschild gravity dual model
respectively.

Introduction of quark masses into the configuration can be done by
terminating hanging strings at certain radial distance $u_{max}<
\infty$.  The universal behaviour of heavy-quark potential comes
from the limit $u_{max}\to \infty$.  We can split the total
binding potential of the string into two parts.  The first part is
the binding potential in the $u_{max}\to \infty$ limit and the
second part is the mass dependent potential.  Therefore, the mass
dependence part of the binding potential, $E_{F1}(u_{max})$~($m =
u_{max}/2 \pi$), can be expressed as
\begin{eqnarray}
E_{\text{F1}}(\text{finite mass)}& = & E_{\text{F1}}(u_{max}\to \infty) + E_{\text{F1}}(u_{max}), \\
E_{\text{F1}}(u_{max})& = & -\frac{u_{c}}{2
\pi}\int_{u_{max}/u{c}}^{\infty}dy \bigg\lbrack \sqrt{\frac{y^n
-x^n}{(y^n-x^n)-(1-x^n)(1-{A(n)}^2)}}-1\bigg\rbrack \\
                      & = & -\frac{u_{max}(1-A(n)^2)}{4 \pi
                      (n-1)}\left( \frac{u_{c}^{n} -
                      u_{T}^{n}}{u_{max}^n}\right)+O(u_{max}^{1-2n}).
\end{eqnarray}
Eliminate $u_{c}$ by using
\begin{equation}\label{qmassl}
 L=\frac{R^{n/2}}{{u_c}^{(n-2)/2}}\int_1^{u_{max}/u_{c}} dy\:
\sqrt{\frac{(1-x^n)(1-A(n)^2)}{(y^n-x^n)(y^n-x^n-(1-x^n)(1-A(n)^2))}}.\\
\end{equation}
The result involves complicated functions of $A$ which can be cast
in the following form,
\begin{eqnarray}
E_{F1}(u_{max})& \sim & -u_{max}^{1-n}\left(
R^{n^{2}/(n-2)}f_{1}(A) + u_{T}^{n}f_{2}(A) \right),
\end{eqnarray}
where $f_{1,2}(A)$ are some functions of $A$.

Interestingly, the mass dependence of multiquark potentials has
similar form as the mass dependence of mesonic state $\sim
m^{1-n}$ in Ref.~\cite{abl}.  This is natural due to the fact that
most of the mass of constituent quarks come from the tail part of
strings which extend to the large-$u$ region.  The mass dependence
of the binding potential at the leading order is therefore
determined only by the contribution of the hanging strings from
the large-$u$ region. As long as the background spacetime of the
gravity dual is asymptotically similar to the background
considered here in the large-$u$ limit, we would expect the same
mass dependence as the form we obtained in this section.

\section{Phase diagram }

A natural question to ask is whether we have a phase where the
exotic multiquark states are preferred over the normal nuclear
matter~(namely the gas of $N$-baryons), vacuum, and the
chiral-symmetric quark-gluon plasma phase. To consider a realistic
model where these three phases are distinct, we focus our
consideration on the Sakai-Sugimoto model with $n=3$.  To
calculate the phase diagram involving exotic states, it is
necessary to consider the contribution from D8 and
$\overline{\text{D8}}$-branes in the Sakai-Sugimoto model in
addition to the contributions from strings and D4-branes.  We will
assume that the characteristic distance between D8 and
$\overline{\text{D8}}$ in $x^{4}$ direction is $L_{0}$. The
relevant scales of the model therefore depend on both $u_{T}$ and
$L_{0}$.

When there is no radial string pulling the vertex down towards the
horizon, it was demonstrated in Ref.~\cite{Guijosa} by numerical
method that the vertex will be pulled all the way up to the
position of the flavour branes if the temperature is not very
high.  Addition of radial strings to the vertex would pull the
vertex and the flavour branes towards the horizon.  As temperature
rises, the radial strings pull the vertex down with stronger force
since they are closer to the horizon.  It is possible that the
vertex then starts to separate from the flavour branes and we
might need to consider the configuration where the vertex and
flavour branes are separated. However, we can see that the
difference between the two configurations should be relatively
small~(namely, only the force conditions will be slightly
different) and we should be able to approximate the situation by
considering the configuration where the vertex is not separated
from the flavour branes.  It is also shown in the Appendix that
this configuration satisfies the force condition and thus is
allowed. Therefore, it will be assumed that the vertex is always
in the flavour branes for the discussion in this section.
Moreover, the vertex will be treated as a static configuration and
any distortion caused by the strings attached to it will be
ignored.

\begin{figure}
\centering
\includegraphics[width=0.8\textwidth]{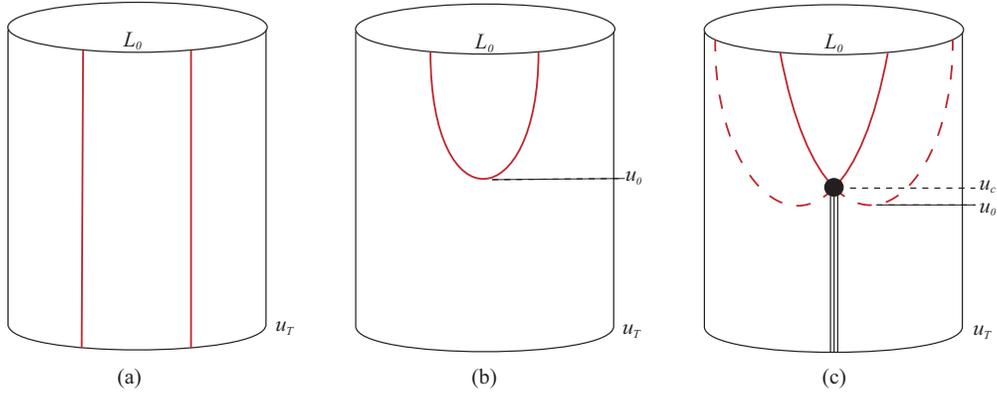}
\caption{Configurations of $\chi$S-QGP~(a), vacuum~(b) and exotic
nuclear phase~(c) in $x^{4}-u$ projection.} \label{phxu}
\end{figure}

The calculations presented in this section are adapted from
Ref.~\cite{bll} except that we add radial strings hanging from the
vertex down to the horizon for the consideration of exotic nuclear
phase.  We also use position of the D4, $u_{c}$, instead of
$u_{0}$~(where $x^{\prime}_{4}(u_{0})\to \infty$) in our
calculation concerning the exotics.  This approach allows us to
deal with the contribution from radial strings more conveniently.
As is shown in Fig.~\ref{phxu}, the vacuum phase with broken
chiral symmetry corresponds to the configuration where D8 and
$\overline{\text{D8}}$ are connected into a curve in the $x_{4}-u$
projection.  The chiral-symmetric phase of quark-gluon
plasma~($\chi$S-QGP) corresponds to the configuration with the
parallel D8 and $\overline{\text{D8}}$ stretching from the
spacetime boundary down to the horizon.  Finally, the
nuclear~(including exotics) phase corresponds to the configuration
where the D4 vertex is located at the D8-$\overline{\text{D8}}$
curve, pulling it down towards the horizon by its ``weight" in the
background.  Each vertex has radial strings attached to it,
pulling it further towards the horizon. When there is no radial
strings attached, the nuclear phase is of normal $N$-baryons.  The
chiral symmetry is also broken in this phase.

Under the above assumptions, the contribution from the strings
hanging down from the spacetime boundary to the vertex is
negligible. The only contribution of strings is from the radial
strings hanging down from the vertex to the horizon.  The total
action of the configuration is given by
\begin{eqnarray}
S & = & S_{D8}+S_{D4}+\tilde{S}_{F1}.
\end{eqnarray}

Generically, the DBI action of the D8-branes is given by
\begin{eqnarray}
S_{D8} & = & -\mu_{8}\int d^{9}X
e^{-\phi}\text{Tr}\sqrt{-det(g_{MN}+2\pi \alpha^{\prime}F_{MN})}
\end{eqnarray}
where the field strength of the flavour group $U(N_{f})$ is
\begin{equation}
{\bf \mathcal{F}} = d\mathcal{A} + i \mathcal{A}\wedge
\mathcal{A}.
\end{equation}
The flavour branes provide ``global" quantum numbers such as
baryon number to the string and subsequently to the strings-brane
configuration dual to baryon in the gauge theory side.  The
diagonal part of the representation matrix of $U(N_{f})$ is the
$U(1)$ subgroup which induces baryon number to the end of string
attached to the flavour branes.  Redefine the $U(1)$ part so that
\begin{eqnarray}
\mathcal{A} & = &
\mathcal{A}_{SU(N_{f})}+\frac{1}{\sqrt{2N_{f}}}\hat{\mathcal{A}}
\end{eqnarray}
with $\hat{\mathcal{A}}$ represents the $U(1)$ piece of the gauge
field. The DBI action of the D8-brane coupled to the diagonal
gauge field is then given by
\begin{eqnarray}
S_{D8}& = & {\mathcal N} \int du ~u^{4}
\sqrt{f(u)(x^{\prime}_{4}(u))^{2}+u^{-3}(1-(\hat{a}^{\prime}_{0}(u))^{2})}
\end{eqnarray}
where the constant scales linearly with $N_{f}$ as
\begin{eqnarray}
{\mathcal N}=\frac{\mu_{8}\tau N_{f}\Omega_{4}V_{3} R^{5}}{g_{s}},
\end{eqnarray}
and the rescaled $U(1)$ diagonal field,
\begin{eqnarray}
\hat{a}& = & \frac{2 \pi
\alpha^{\prime}\hat{\mathcal{A}}}{R\sqrt{2 N_{f}}}.
\end{eqnarray}
The action does not depend on $\hat{a}_{0}(u)$ explicitly and
therefore a constant of motion can be defined as
\begin{eqnarray}
d& = & \frac{u
\hat{a}^{\prime}_{0}(u)}{\sqrt{f(u)(x^{\prime}_{4}(u))^{2}+u^{-3}(1-(\hat{a}^{\prime}_{0}(u))^{2})}}.
\end{eqnarray}
We will see below that the constant $d$ can be interpreted as the
baryon number density sourced by the D4-branes once we introduce
the Chern-Simon action of the gauge field.  Note that $d$ plays
the role of the electric displacement field \cite{bll}.  In the
confined phase, the only possible source for $d$ is the D4-brane
wrapped on $S^4$ in the D8-branes.  In the deconfined phase,
either D4-brane or strings which stretch from the D8-brane down to
the horizon can serve as the sources for $d$.  Here, in the study
of exotic baryons, we consider the case where {\it both} D4-brane
and strings are present as the sources. This possibility was not
investigated in \cite{bll}.

Similarly, the constant of motion with respect to $x_{4}(u)$ leads
to
\begin{eqnarray}
(x^{\prime}_{4}(u))^{2}& = & \frac{1}{u^{3}f(u)}\Big[
\frac{f(u)(u^{8}+u^{3}d^{2})}{f(u_{0})(u^{8}_{0}+u^{3}_{0}d^{2})}-1
\Big]^{-1}
\end{eqnarray}
where $u_{0}$ is the position when $x^{\prime}_{4}(u_{0})=\infty$.

Instead of using $u_{0}$ as the reference position, the radial
position of the D4 on the D8-branes, $u_{c}$, can be used to
calculate $x^{\prime}_{4}(u)$,
\begin{eqnarray}
(x^{\prime}_{4}(u))^{2}& = & \frac{1}{u^{3}f(u)}\Big[
\frac{f(u)(u^{8}+u^{3}d^{2})}{F^{2}}-1 \Big]^{-1}
\end{eqnarray}
where
\begin{eqnarray}
F& = & \frac{f(u_{c})
\sqrt{u_{c}^{8}+u_{c}^{3}d^2}}{\sqrt{f(u_{c})(x^{\prime}_{4}(u_{c}))^{2}+u_{c}^{-3}}}x^{\prime}_{4}(u_{c})
\\
 & = & \frac{\sqrt{u_{c}^{3}f(u_{c})}}{3}\Big[ 1+\frac{1}{2}\left(
\frac{u_{T}}{u_{c}}\right)^{3}+3 n_{s}\sqrt{f(u_{c})} \Big]\sqrt{
 \frac{9(u_{c}^{5}+d^{2})}{1+\frac{1}{2}(\frac{u_{T}}{u_{c}})^{3}+3
 n_{s}\sqrt{f(u_{c})}}-\frac{d^{2}}{f(u_{c})}}.
\end{eqnarray}
The number of radial strings $n_{s}$ represents the number of
strings hanging down from D4-branes to the horizon in unit of
$1/N$.  For $k,(N+\bar{k})$-baryon and $j$-mesonance, the values
of $n_{s}$ are $1-k/N,\bar{k}/N,1$ respectively. Calculation of
$x^{\prime}_{4}(u_{c})$ is performed by minimizing the action with
respect to the variation of $u_{c}$~(see Appendix).  For a fixed
$L_{0}$, increasing the number of strings $n_s$ results in D4-D8
configuration being pulled down more towards the horizon.

The $U(N_{f})$ gauge field $\mathcal{A}$ also generates
Chern-Simon term,
\begin{eqnarray}
S_{CS} & = & \frac{N}{24 \pi^{2}}\int_{M^{4}\times
R}\omega_{5}(\mathcal{A}).
\end{eqnarray}
For $\mathcal{A} = \mathcal{A}_{\mu}dx^{\mu}+\mathcal{A}_{u}du$,
the 5-form field strength is given by
\begin{eqnarray}
\omega_{5}(\mathcal{A}) & = & Tr\left( \mathcal{A}{\bf
\mathcal{F}}^{2}-\frac{1}{2}\mathcal{A}^{3}{\bf
\mathcal{F}}+\frac{1}{10}\mathcal{A}^{5} \right).
\end{eqnarray}
Only the first term contains non-vanishing contribution from the
$U(1)$ part which would be identified with the number density of
baryon.  We will assume a uniform distribution $n_{4}$ of the gas
of D4-branes in ${\mathbb R}^{3}$ at $u=u_{c}$ in the radial
direction.  This leads to the relation between the number density
of D4-branes, $n_{4}$, and baryon number density $d$~\cite{bll},
\begin{eqnarray}
n_{4}& = & \frac{2 \pi \alpha^{\prime} R^{2} {\mathcal N}}{\tau
V_{3}N}d.
\end{eqnarray}

Phase transition for a system where the number of particles varies
is most conveniently described by the grand canonical ensemble.
The grand canonical potential of each phase can be defined using
the corresponding action of the D8-branes as
\begin{eqnarray}
\Omega(\mu) & = & \frac{1}{{\mathcal
N}}S_{D8}[x_{4}(u),\hat{a}_{0}(u)]_{cl}.
\end{eqnarray}
The baryon chemical potential is given by the $U(1)$ diagonal
field at the boundary,
\begin{eqnarray}
\mu & = & \hat{a}_{0}(\infty),
\end{eqnarray}
from which the baryon number density is determined,
\begin{eqnarray}
d & = & -\frac{\partial \Omega(\mu)}{\partial \mu}.
\end{eqnarray}
This justifies the association of grand canonical potential with
the D8 action.  When additional sources of the baryon number are
introduced, the free energy, ${\mathcal F}_{E}$, from the sources
will also contribute to the baryon chemical potential,
\begin{eqnarray}
\mu & = & \frac{\partial}{\partial d}\frac{1}{{\mathcal N}}\left(
\tilde{S}_{D8}[x_{4}(u),d(u)]_{cl}+S_{source}(d,u_{c})
\right)\equiv \frac{\partial {\mathcal F}_{E}}{\partial d},
\end{eqnarray}
where the Legendre-transformed action $\tilde{S}_{D8}$ is given by
\begin{eqnarray}
\tilde{S}_{D8} & = & S_{D8} + {\mathcal N}
\int^{\infty}_{u_{c}}d(u)\hat{a}^{\prime}_{0}~du, \\
               & = & {\mathcal N} \int^{\infty}_{u_{c}} du~ u^{4}
\sqrt{f(u)(x^{\prime}_{4}(u))^{2}+u^{-3}}\sqrt{1+\frac{d^{2}}{u^{5}}}.
\end{eqnarray}
In our case, the additional sources are D4 and radial strings.
These relations can also be applied to the vacuum phase~(with
broken chiral symmetry) where $u_{c}$ is replaced with $u_{0}$.

Setting $L_{0}=2\int^{\infty}_{u_{i}=u_{0},u_{c}}
x^{\prime}_{4}(u)du=1$, the expressions for the grand canonical
potential and the chemical potential for each phase are given by

\underline{vacuum phase, $d=0$}:
\begin{eqnarray}
\Omega_{vac} & = & \int^{\infty}_{u_{0}}du
\frac{u^{5/2}\sqrt{f(u)}}{\sqrt{f(u)-
\frac{u^{8}_{0}}{u^{8}}f(u_{0})}},
\end{eqnarray}

\underline{$\chi$S-QGP phase, $x^{\prime}_{4}(u)=0$}:
\begin{eqnarray}
\Omega_{qgp} & = & \int^{\infty}_{u_{T}}du
\frac{u^{5}}{\sqrt{u^{5}+d^{2}}}, \\
\mu_{qgp} & = & \int^{\infty}_{u_{T}}du
\frac{d}{\sqrt{u^{5}+d^{2}}},
\end{eqnarray}

\underline{nuclear~(including exotics) phase }:
\begin{eqnarray}
\Omega_{nuc} & = & \int^{\infty}_{u_{c}}du~
\Big[1-\displaystyle{\frac{F^{2}}{f(u)(u^{8}+u^{3}d^{2})}}\Big]
^{-1/2}
\frac{u^{5}}{\sqrt{u^{5}+d^{2}}}, \\
\mu_{nuc} & = & \int^{\infty}_{u_{c}}du~
\Big[1-\displaystyle{\frac{F^{2}}{f(u)(u^{8}+u^{3}d^{2})}}\Big]^{-1/2}
\frac{d}{\sqrt{u^{5}+d^{2}}} +
\frac{1}{3}u_{c}\sqrt{f(u_{c})}+n_{s}(u_{c}-u_{T}).
\end{eqnarray}
At a fixed temperature $T$ and chemical potential $\mu$, a first
order phase transition line between phase 1 and 2 is obtained when
$\Omega_{1}=\Omega_{2},\mu_{1}=\mu_{2}=\mu$.  Transitions between
vacuum $\leftrightarrow$ $\chi$S-QGP and $\chi$S-QGP
$\leftrightarrow$ nuclear phases are of this kind.  On the other
hand, phase transition between nuclear $\leftrightarrow$ vacuum is
second order in nature, at least for this case when there is no
interaction between each D4.  The second order phase transition
line occurs when
\begin{eqnarray}
\frac{\partial \mu}{\partial d}& = & \frac{\partial^{2} {\mathcal
F}_{E}}{\partial d^{2}}
\end{eqnarray}
has discontinuity at $d=0$.

\begin{figure}
\centering
\begingroup
  \makeatletter
  \providecommand\color[2][]{%
    \GenericError{(gnuplot) \space\space\space\@spaces}{%
      Package color not loaded in conjunction with
      terminal option `colourtext'%
    }{See the gnuplot documentation for explanation.%
    }{Either use 'blacktext' in gnuplot or load the package
      color.sty in LaTeX.}%
    \renewcommand\color[2][]{}%
  }%
  \providecommand\includegraphics[2][]{%
    \GenericError{(gnuplot) \space\space\space\@spaces}{%
      Package graphicx or graphics not loaded%
    }{See the gnuplot documentation for explanation.%
    }{The gnuplot epslatex terminal needs graphicx.sty or graphics.sty.}%
    \renewcommand\includegraphics[2][]{}%
  }%
  \providecommand\rotatebox[2]{#2}%
  \@ifundefined{ifGPcolor}{%
    \newif\ifGPcolor
    \GPcolorfalse
  }{}%
  \@ifundefined{ifGPblacktext}{%
    \newif\ifGPblacktext
    \GPblacktexttrue
  }{}%
  % define a \g@addto@macro without @ in the name:
  \let\gplgaddtomacro\g@addto@macro
  % define empty templates for all commands taking text:
  \gdef\gplbacktext{}%
  \gdef\gplfronttext{}%
  \makeatother
  \ifGPblacktext
    % no textcolor at all
    \def\colorrgb#1{}%
    \def\colorgray#1{}%
  \else
    % gray or color?
    \ifGPcolor
      \def\colorrgb#1{\color[rgb]{#1}}%
      \def\colorgray#1{\color[gray]{#1}}%
      \expandafter\def\csname LTw\endcsname{\color{white}}%
      \expandafter\def\csname LTb\endcsname{\color{black}}%
      \expandafter\def\csname LTa\endcsname{\color{black}}%
      \expandafter\def\csname LT0\endcsname{\color[rgb]{1,0,0}}%
      \expandafter\def\csname LT1\endcsname{\color[rgb]{0,1,0}}%
      \expandafter\def\csname LT2\endcsname{\color[rgb]{0,0,1}}%
      \expandafter\def\csname LT3\endcsname{\color[rgb]{1,0,1}}%
      \expandafter\def\csname LT4\endcsname{\color[rgb]{0,1,1}}%
      \expandafter\def\csname LT5\endcsname{\color[rgb]{1,1,0}}%
      \expandafter\def\csname LT6\endcsname{\color[rgb]{0,0,0}}%
      \expandafter\def\csname LT7\endcsname{\color[rgb]{1,0.3,0}}%
      \expandafter\def\csname LT8\endcsname{\color[rgb]{0.5,0.5,0.5}}%
    \else
      % gray
      \def\colorrgb#1{\color{black}}%
      \def\colorgray#1{\color[gray]{#1}}%
      \expandafter\def\csname LTw\endcsname{\color{white}}%
      \expandafter\def\csname LTb\endcsname{\color{black}}%
      \expandafter\def\csname LTa\endcsname{\color{black}}%
      \expandafter\def\csname LT0\endcsname{\color{black}}%
      \expandafter\def\csname LT1\endcsname{\color{black}}%
      \expandafter\def\csname LT2\endcsname{\color{black}}%
      \expandafter\def\csname LT3\endcsname{\color{black}}%
      \expandafter\def\csname LT4\endcsname{\color{black}}%
      \expandafter\def\csname LT5\endcsname{\color{black}}%
      \expandafter\def\csname LT6\endcsname{\color{black}}%
      \expandafter\def\csname LT7\endcsname{\color{black}}%
      \expandafter\def\csname LT8\endcsname{\color{black}}%
    \fi
  \fi
  \setlength{\unitlength}{0.0500bp}%
  \begin{picture}(7200.00,5040.00)%
    \gplgaddtomacro\gplbacktext{%
      \csname LTb\endcsname%
      \put(1166,660){\makebox(0,0)[r]{\strut{}$0.025$}}%
      \put(1166,1265){\makebox(0,0)[r]{\strut{}$0.050$}}%
      \put(1166,1871){\makebox(0,0)[r]{\strut{}$0.075$}}%
      \put(1166,2476){\makebox(0,0)[r]{\strut{}$0.100$}}%
      \put(1166,3081){\makebox(0,0)[r]{\strut{}$0.125$}}%
      \put(1166,3686){\makebox(0,0)[r]{\strut{}$0.150$}}%
      \put(1166,4292){\makebox(0,0)[r]{\strut{}$0.175$}}%
      \put(1300,440){\makebox(0,0){\strut{}$  0.02$}}%
      \put(2730,440){\makebox(0,0){\strut{}$  0.1$}}%
      \put(4778,440){\makebox(0,0){\strut{}$  1.0$}}%
      \put(6826,440){\makebox(0,0){\strut{}$ 10.0$}}%
      \put(264,2760){{\makebox(0,0){$T$}}}%
      \put(4062,110){\makebox(0,0){$\mu$}}%

    }%
    \gplgaddtomacro\gplfronttext{%
    }%
    \gplbacktext
    \put(1295,655){\includegraphics{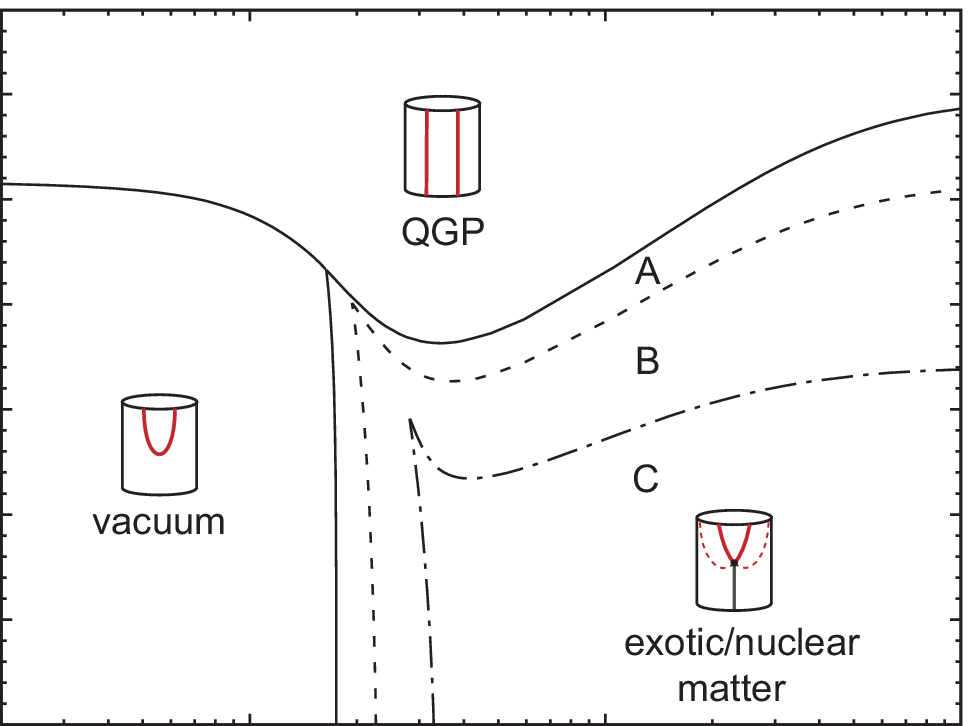}}%
    \gplfronttext
  \end{picture}%
\endgroup
\caption{The phase diagram of exotic nuclear matters above the
deconfinement temperature. Nuclear phase including exotics is
shown as the region on the lower right corner where it is divided
into 3 parts for representative purpose.  $A,B,C$ represents the
region where exotic baryon phase with
$n_{s}=0~(N\text{-baryon}),0.1,0.3$ is preferred over vacuum and
$\chi$S-QGP respectively.} \label{phdt}
\end{figure}

In the Sakai-Sugimoto model, there is a phase transition
temperature above which gluons become deconfined.  However, it
does not necessarily imply that everything including quark and
antiquark is totally free and chiral symmetry is completely
restored above this temperature. When the baryon chemical
potential is sufficiently high, baryons can exist even when the
temperature is higher than the deconfinement
temperature~\cite{bll}.  Only when the temperature increases even
further that everything will be completely dissolved and the
chiral symmetry is also restored.  We also see this behavior in
the phase diagram in Figure~\ref{phdt} where we ignore the
confined region at low temperature and present only the deconfined
part of the phase diagram.

The phase diagram of vacuum with broken chiral symmetry,
$\chi$S-QGP and phase of nuclear including exotic multi-quark
states is shown in Figure~\ref{phdt}.  The phase diagram involving
vacuum and $\chi$S-QGP phases was first obtained in
Ref.~\cite{Tanii} and the full phase diagram without the exotics
was obtained in Ref.~\cite{bll}.  Since the strings pull down the
D4-D8 configuration towards the horizon, the configuration with
$n_{s}>0$ is less stable than the normal $N$-baryon~($n_{s}=0$).
This is shown in Fig.~\ref{phdt} where the region of $n_{s}>0$
nuclear phase~($B,C$) is smaller than the region of $N$-baryon
phase~($A$).  They are actually less stable than the $N$-baryon
since the grand canonical potential
$\Omega_{n_{s}>0}(T,\mu)>\Omega_{n_{s}=0}(T,\mu)$ for $0.5>
n_{s}>0$.  Above $n_{s}>0.3$, the exotic phase becomes unstable to
density fluctuations~($\frac{\partial \mu}{\partial d}<0$) at high
temperatures in certain range of $d$ but still remains stable in a
region of parameter space. Numerical studies reveal that for
approximately $n_{s}> 0.5$, the multiquark states become unstable
thermodynamically with respect to density fluctuations for most of
the temperatures.

Addition of radial strings introduces extra source of the baryonic
chemical potential.  We can see from Fig.~\ref{phdt} that the
value of $\mu_{onset}$ for the exotic nuclear phase increases with
the value of $n_{s}$.  Nevertheless, once emerged~(i.e. $\mu >
\mu_{onset}$), the exotic phases are more stable than the vacuum
at any temperature, but less stable than $\chi$S-QGP at
sufficiently high temperatures above which chiral symmetry is
restored.

\section{Discussions}

It is desirable to compare the binding energy of each multi-quark
state in order to discuss the stability of each configuration as
well as their relative abundances in the deconfined phase. At a
fixed temperature $T$, we can compare numerically the binding
energies $E$ as functions of the size $L$ of the configuration as
is shown in Fig.~\ref{ceb},\ref{cem}.  For $k$-baryon and
$(N+\bar{k})$-baryon, we compare the energy with $N$-baryon.  For
$j$-mesonance, we compare the energy with the energy of $j$
mesons.

From Fig.~\ref{ceb}, $N$-baryon is more energetically favoured
than $k$-baryon and $(N+\bar{k})$-baryon for any value of $k,
\bar{k}$. Since there are less hanging strings from the spacetime
boundary and more radial strings pulled down into the horizon in
the case of $k$-baryon, the vertex is located closer to the
horizon and consequently becomes less energetically favoured
comparing to the $N$-baryon.  Similarly in the case of
$(N+\bar{k})$, even though not as obvious, adding $\bar{k}$
hanging and radial strings to the configuration of $N$-baryon
results in positive energy increase in the binding potential,
making this configuration less favoured energetically.  An
$(N+\bar{k})$-baryon naturally tends to decay into $N$-baryon plus
$\bar{k}$ free antiquark strings. A $k$-baryon also has the
tendency to fuse with $(N-k)$ quarks to form an $N$-baryon with
lower energy.

The situation of $j$-mesonance is somewhat similar.  Even though
$j$ mesons are always energetically preferred over $j$-mesonance
for all values of $j$, $j$-mesonance with higher value of $j$ has
stronger binding force than the lower ones as is shown in
Fig.~\ref{cem}.  From the energy viewpoint, $j$-mesonance will
prefer to split into a number of $j$ mesons.  It is notable that
the screening length of $j$-mesonance will approach the value of
meson, $L^{*}_{meson}$, but it will never exceed $L^{*}_{meson}$.

For the case of $(N+\bar{k})$-baryon and $j$-mesonance, there
exist the limits $\bar{k}\to \infty$ and $j \to \infty$.  The
first limit for $(N+\bar{k})$-baryon leads to the zero-size
configuration which saturates the zero-force condition. The second
limit for $j$-mesonance leads to the {\it mesonic} limit where the
configuration is similar to the system of $j$ mesons as we will
see in the following.

From Eqn.~(\ref{An}), since $A(n)\sim ({j/N})^{-1}$, $A(n)$
becomes negligible for large $j/N$.  Therefore, we can neglect
$A(n)$ and obtain that $E_{F1}$ does not depend on $j/N$. Using
asymptotic expansions, Eqn.~\eqref{mathcalE} becomes
\begin{align}
\mathcal{E} &\simeq \Bigg\lbrace \int_1^\infty dy \bigg[
\sqrt{\frac{y^n-x^n}{y^n-1}}-1\bigg]
-(1-x)\Bigg\rbrace\nonumber\\
       &= \Bigg\{ u_T - \frac{\Gamma \left(\frac{1}{2}\right)
       \Gamma \left(1-\frac{1}{n}\right)}{\Gamma \left(\frac{1}{2} -\frac{1}{n}\right)}\frac{
       C^{2/(n-2)}}{L^{2/(n-2)}}\Bigg\}+\mathcal{O} (x^n),
\end{align}
where
\begin{equation}
C(n) \equiv
\frac{R^{n/2}}{n}\frac{\Gamma(1-\frac{1}{n})\Gamma(\frac{1}{2})}{\Gamma(\frac{3}{2}-\frac{1}{n})}.
\nonumber
\end{equation}

Now, consider Eqn.~\eqref{Etotrj}, we find the screening length
$L_{*}$~(half the distance between quarks at which the binding
energy is zero) by setting $E_{tot}=0$. In the limit of $j/N$
becoming very large, we can obtain $L_{*}$ from the condition

\begin{equation}\label{findL*}
\mathcal{E}(L_{*})=0,
\end{equation}
leading to
\begin{equation}
L_{*}\simeq \bigg[\frac{\Gamma (\frac{1}{2})\Gamma
(1-\frac{1}{n})}{u_T (n)\Gamma
(\frac{1}{2}-\frac{1}{n})}\bigg]^{(n-2)/2} C(n).
\end{equation}
Again, the case $n=3$ and $n=4$ correspond to the Sakai-Sugimoto
and the AdS-Schwarzschild gravity dual model respectively.  This
expression is exactly the same as the screening length of meson in
the deconfined phase from Ref.~\cite{abl}\footnote{Our definition
of the screening length is one-half of the definition in
Ref.~\cite{abl}.}. It is no surprise since in the $j \to \infty$
limit, the hanging strings from the boundary exert force
overwhelmingly, therefore the ``weight" of the baryon vertex plus
the tension of radial strings become negligible. Effectively, the
end of hanging string at the vertex will feel zero force down and
thus the slope $u'_{c}$ will be zero.  As a result, the strings
from the boundary will hang smoothly and appear similar to hanging
strings in the case of the mesonic state.

Even in the deconfined phase, we therefore perceive that in
addition to free quarks, antiquarks, and gluons, there will also
be mesons and multi-quark states. Due to the lower energy, there
are more $N$-baryons than $(N+\bar{k})$-baryons and $k$-baryons.
The relative populations can be estimated using the Boltzmann
factor
\begin{eqnarray}
\text{exp}(-\frac{E}{k_{B} T}),
\end{eqnarray}
determined by the corresponding binding energy $E$ for each state.

A more precise way of considering the deconfined phase is to use
the grand canonical potential as the indicator for the stable
phase. Following Bergman, Lifschytz, and Lippert~\cite{bll}, we
consider three phases of the deconfined soup, a vacuum phase and a
nuclear phase with broken chiral symmetry, and a $\chi$S-QGP. For
sufficiently high chemical potential and moderate temperature, the
nuclear phase of the multiquark states is preferred over the
vacuum and $\chi$S-QGP phase. Exotic nuclear states such as
$k$-baryon, $(N+\bar{k})$-baryon, and $j$-mesonance are
characterized by the number of radial strings $n_{s}$ hanging down
from the D4-branes to the horizon. It is found that the multiquark
states with $n_{s}>0.5$ are unstable thermodynamically.  However,
all of these exotic states with $0.5\geq n_{s}>0$ are less stable
than the normal $N$-baryon with $n_{s}=0$.

For each value of $n_{s}$, there exists a triple point where the
grand canonical potentials of the three phases are equivalent.
Varying $n_{s}$, this triple point will move along the phase
transition line between the vacuum and the $\chi$S-QGP as is shown
in Fig.~\ref{phdt}. The stable region of the nuclear phase shrinks
as $n_{s}$ increases. As $n_{s}>0.5$, the nuclear phase becomes
thermodynamically unstable with respect to the density
fluctuations for most of the parameter space.

\section{Conclusion}

The gravity dual picture of the deconfined phase suggests that the
binding energy or potential between quarks and antiquarks in this
phase is nonzero due to the Coulombic piece of the interaction.
Since the colorless condition is not required in the deconfined
phase, exotic configurations of the multiquark states are
possible. We investigate three classes of these configurations,
$k$-baryon, $(N+\bar{k})$-baryon, and $j$-mesonance.  It is found
that all of these configurations are less energetically favoured
than the normal $N$-baryon as well as being less stable
thermodynamically.

The dependence of the screening length on the parameters
$k,\bar{k},j$ is studied and the results are shown in
Fig.~\ref{slk}-\ref{slj}. The screening length of $k$-baryon and
$j$-mesonance are notably increasing with the values of $k$ and
$j$ whereas the screening length of $(N+\bar{k})$-baryon is a
decreasing function of $\bar{k}$. Interestingly, $j$-mesonance has
saturated value of screening length equal to the screening length
of meson as $j\to \infty$.

The dependence on the quark mass of the binding potential at the
leading order is derived and found to be $\sim m^{1-n}$~($n=3,4$
for the Sakai-Sugimoto, AdS-Schwarzschild model).  The linear
quark-mass dependence of the rest energy that we naturally expect
is included in the regulator and therefore not present in the
binding potential.

In order to consider phase diagram involving exotic nuclear phase,
we consider the Sakai-Sugimoto model where the flavour branes D8
and $\overline{\text{D8}}$ are introduced.  The flavour D8-branes
action is identified with the grand canonical potential of the
relevant phase.  The nuclear phase is considered in the limit when
the D4-branes are pulled all the way up to the flavour branes.
Exotic multiquark states with a number of strings stretched down
to the horizon, i.e. $n_{s}>0$, become less stable than normal
$N$-baryon~($n_{s}=0$) since radial strings attached to the
D4-branes pull the D4-D8 configuration closer to the horizon.
Nevertheless, comparing to the vacuum and the $\chi$S-QGP phase,
the nuclear phase of exotic multiquark states can be more stable
in a region of phase diagram with high chemical potential and low
temperature as is shown in Fig.~\ref{phdt}.  In this region, we
expect to have a nuclear phase where $N$-baryons, $k$-baryons, and
$(N+\bar{k})$-baryons coexist. For $j$-mesonance with $n_{s}=1$,
our consideration of the grand canonical potential suggests that
it is thermodynamically unstable to density fluctuations since
$\frac{\partial \mu}{\partial d}<0$. Generically, numerical
studies reveal that exotic baryons with $n_{s}>0.5$~(namely
$k$-baryon with $k/N<0.5$, $(N+\bar{k})$-baryon with
$\bar{k}/N>0.5$ and any $j$-mesonance) in the deconfined phase are
thermodynamically unstable to density fluctuations.

\section*{Acknowledgments}
\indent We would like to thank Wen-Yu Wen, Ahpisit Ungkitchanukit
and Kazuyuki Furuuchi for valuable comments.  E.H. is supported by
the Commission on Higher Education~(CHE), Thailand under the
program Strategic Scholarships for Frontier Research Network for
the Ph.D. Program Thai Doctoral Degree for this research.  P.B.
and A.C. is supported in part by the Thailand Research Fund~(TRF)
and Commission on Higher Education~(CHE) under grant MRG5180227
and MRG5180225 respectively.

\appendix
\section{Force condition at the D8-branes}

There are three forces acting on a D4 locating inside the
D8-branes, one from the D8, another from the radial strings
pulling down towards horizon and lastly the force from its own
``weight" in the background.  The equilibrium can be sustained
only when these three forces are balanced.  As is shown in
Ref.~\cite{bll}, variation of the total action with respect to
$u_{c}$ and the constant of motion with respect to $x_{4}(u)$ lead
to
\begin{eqnarray}
x^{\prime}_{4}(u_{c})& = & \displaystyle{ \left( \tilde{L}(u_{c})
-
\frac{\partial{S_{source}}}{\partial{u_{c}}}\right)\Bigg{/}{\frac{\partial
\tilde{S}_{D8}}{\partial{x^{\prime}_{4}}}\bigg{\vert}_{u_{c}}}}, \\
                     & = & \frac{1}{d}\sqrt{
 \frac{9u_{c}^{2}(1+\frac{d^{2}}{u_{c}^{5}})}{1+\frac{1}{2}(\frac{u_{T}}{u_{c}})^{3}+3
 n_{s}\sqrt{f(u_{c})}}-\frac{d^{2}u_{c}^{-3}}{f(u_{c})}}
\end{eqnarray}
where the Legendre transformed action is
\begin{eqnarray}
\tilde{S}_{D8} & = &
\int^{\infty}_{u_{c}}\tilde{L}(x^{\prime}_{4}(u),d)\,du, \\
               & = & {\mathcal N} \int^{\infty}_{u_{c}} du~ u^{4}
\sqrt{f(u)(x^{\prime}_{4}(u))^{2}+u^{-3}}\sqrt{1+\frac{d^{2}}{u^{5}}},
\end{eqnarray}
and the source term is given by
\begin{eqnarray}
S_{source}     & = & {\mathcal N} d \Big[
\frac{1}{3}u_{c}\sqrt{f(u_{c})}+n_{s}(u_{c}-u_{T})\Big].
\end{eqnarray}
There are two contributions from the D-branes and strings as the
sources for the baryon chemical potential.  Additional strings
increase the baryonic chemical potential of the exotic multiquark
states. Since the number of total charge on each D4 is $N$ which
is absorbed into $\mathcal{N}$, the number of radial strings
stretched down to the horizon, $n_{s}$, is thus given in unit of
$1/N$.


\newpage

\begin{thebibliography}{11}

\bibitem{maldacena}
J. M. Maldacena, ``The Large N Limit of Superconformal Field
Theories and Supergravity,'' {\it Adv. Theor. Math. Phys.} {\bf 2}
(1998) 231-252 [{\it Int. J. Theor. Phys.} {\bf 38} (1998)
1113-1133], [arXiv:hep-th/9711200].
\bibitem{agmoo}
O. Aharony, S. S. Gubser, J. M. Maldacena, H. Ooguri, Y. Oz,
``Large N Field Theories, String Theory and Gravity,'' {\it Phys.
Rept.} {\bf 323} (2000) 183-386, [arXiv:hep-th/9905111].
\bibitem{witb}
E. Witten, ``Baryons and Branes in Anti-de Sitter Space,'' {\it
JHEP} {\bf 07} (1998) 006, [arXiv:hep-th/9805112].
\bibitem{gross&ooguri}
D. J. Gross and H. Ooguri, ``Aspects of large N gauge theory
dynamics as seen by string theory,'' {\it Phys. Rev.} {\bf D58}
(1998) 106002, [arXiv:hep-th/9805129].
\bibitem{Imamura}
Y. Imamura, ``String Junctions and Their Duals in Heterotic String
Theory,'' {\it Prog. Theor. Phys.} 101 (1999) 1155,
[arXiv:hep-th/9901001].
\bibitem{Callan}
C. G. Callan, A. Guijosa and K. G. Savvidy, ``Baryons and String
Creation from the Fivebrane Worldvolume Action,'' {\it Nucl.
Phys.} B 547 (1999) 127 [arXiv:hepth/9810092].
\bibitem{Guijosa}
C. G. Callan, A. Guijosa, K. G. Savvidy and O. Tafjord, ``Baryons
and Flux Tubes in Confining Gauge Theories from Brane Actions,''
{\it Nucl. Phys.} B 555 (1999) 183 [arXiv:hep-th/9902197].
\bibitem{BISY}
A. Brandhuber, N. Itzhaki, J. Sonnenschein and S. Yankielowicz, ``Baryon from supergravity,'' {\it JHEP} {\bf 07} (1998) 046, [arXiv:hep-th/9806158].
\bibitem{gho_multi-q}
K. Ghoroku, M. Ishihara, A. Nakamura and F. Toyoda, ``Multi-Quark Baryons and Color Screening at Finite Temperature,'' [arXiv:0806.0195 [hep-th]].
\bibitem{gho_k-quark}
K. Ghoroku and M. Ishihara, ``Baryons with D5 Brane Vertex and $k$-Quarks States,'' {\it Phys. Rev.} {\bf D77} (2008) 086003, [arXiv:0801.4216 [hep-th]].
\bibitem{Car}
M.V. Carlucci, F. Giannuzzi, G. Nardulli, M. Pellicoro and S. Stramaglia,  ``AdS-QCD quark-antiquark potential, meson spectrum and tetraquarks,'' arXiv:0711.2014 [hep-ph].
\bibitem{Wen}
W-Y. Wen, ``Multi-quark potential from AdS/QCD,'' arXiv:0708.2123 [hep-th].
\bibitem{ss lowE}
T. Sakai and S. Sugimoto, ``Low Energy Hadron Physics in Holographic QCD,'' {\it Prog. Theor. Phys.} {\bf 113} (2005) 843, [arXiv:hep-th/0412141].
\bibitem{ss more}
T. Sakai and S. Sugimoto, ``More on a Holographic Dual of QCD,'' {\it Prog. Theor. Phys.} {\bf 114} (2005) 1083, [arXiv:hep-th/0507073].
\bibitem{Wittcon}
E. Witten, ``Anti-de Sitter Space, Thermal Phase Transition, and
Confinement in Gauge Theories,'' {\it Adv. Theor. Math. Phys.} 2
(1998) 505 [arXiv:hep-th/9803131].
\bibitem{Aharony}
O. Aharony, J. Sonnenschein and S. Yankielowicz, ``A Holographic Model of Deconfinement and Chiral Symmetry Restoration,'' {\it Annals Phys.} {\bf 322} (2007) 1420, [arXiv:hep-th/0604161].
\bibitem{Kim}
K. Y. Kim, S. J. Sin and I. Zahed, ``Dense hadronic matter in holographic QCD,'' arXiv:hep-th/0608046.
\bibitem{Tanii}
N. Horigome and Y. Tanii, ``Holographic chiral phase transition with chemical potential,''  {\it JHEP} {\bf 01} (2007) 072, [arXiv:hep-th/0608198].
\bibitem{bll}
Oren Bergman, Gilad Lifschytz, Matthew Lippert, ``Holographic Nuclear Physics," {\it JHEP} {\bf 11} (2007) 056, [arXiv:hep-th/0708.0326].
\bibitem{abl}
O. Antipin, P. Burikham and J. Li, `` Effective Quark Antiquark
Potential in the Quark-Gluon Plasma from Gravity Dual Models,''
{\it JHEP} {\bf 06} (2007) 046, [arXiv:hep-ph/0703105].
\bibitem{bl}
Piyabut Burikham and Jun Li, ``Aspects of the Screening Length and
Drag Force in Two Alternative Gravity Duals of the Quark-Gluon
Plasma," {\it JHEP} {\bf 03} (2007) 067, [arXiv:hep-ph/0701259].

\end{thebibliography}
\end{document}